\documentclass[12pt]{article}
\usepackage{amsmath, amssymb}
\usepackage{graphics,graphicx}
\usepackage{epsfig,comment}

\newcommand{\nn}{\nonumber}
\newcommand{\half}{\frac{1}{2}}

\newcommand{\identity}{{\rlap{1} \hskip 1.6pt \hbox{1}}}
\newcommand{\lsim}{\,\raise.3ex\hbox{$<$\kern-.75em\lower1ex\hbox{$\sim$}}\,}
\newcommand{\gsim}{\,\raise.3ex\hbox{$>$\kern-.75em\lower1ex\hbox{$\sim$}}\,}

\newcommand{\LL}{\mathcal{L}}

\newcommand{\NN}{\mathcal{N}}
\newcommand{\OO}{\mathcal{O}}

\newcommand{\phys}{{\text{phys}}}
\newcommand{\LE}{{\text{LE}}}
\newcommand{\HE}{{\text{HE}}}
\newcommand{\diag}{\text{diag}}
\newcommand{\Tr}{\text{ Tr }}
\newcommand{\TeV}{\text{ TeV }}
\newcommand{\GeV}{\text{ GeV}}
\newcommand{\SM}{{\text{SM}}}
\newcommand{\eff}{{\text{eff}}}
\newcommand{\PGB}{{\text{PGB}}}

\newcommand{\Qc}{{Q^c}}

\begin{document}

\begin{titlepage}
\renewcommand{\thefootnote}{\fnsymbol{footnote}}
\setcounter{footnote}{0}
\begin{flushright}
SLAC-PUB-10701\\
hep-ph/0409127\\
\end{flushright}
\vskip 2cm
\begin{center}
{\large\bf D-Terms, Unification, and the Higgs Mass}
\vskip 1cm
{\normalsize
Alexander Maloney$^{1,2}$, Aaron Pierce$^{1,2}$\footnote{The work of AM and 
AP is supported by the U.S. Department of Energy under contract number DE-AC02-76SF00515.}, 
Jay G. Wacker$^{2}$\footnote{JGW is supported by National Science 
Foundation Grant PHY-9870115 and by the Stanford Institute for Theoretical Physics.}\\
\vskip 0.5cm
1. Theory Group \\
   Stanford Linear Accelerator Center\\
   Menlo Park, CA 94025\\
\vskip .1in
2. Institute for Theoretical Physics\\
   Stanford University\\
   Stanford, CA 94305\\
\vskip .1in
}
\end{center}

\vskip .5cm

\begin{abstract}
We study gauge extensions of the MSSM that contain non-decoupling
$D$-terms, which contribute to the Higgs boson mass.
These models naturally maintain gauge coupling unification and 
raise the Higgs mass without fine-tuning.
Unification constrains the structure of the gauge extensions, 
limiting the Higgs 
mass in these models to $m_{h} \lsim \mbox{150 $\GeV$}$.
The $D$-terms contribute to the Higgs mass only if the extended
gauge symmetry is broken at energies of a few TeV, leading
to new heavy gauge bosons in this mass range.
\end{abstract}
 
\end{titlepage}

\tableofcontents

\section{Introduction}
\renewcommand{\thefootnote}{\arabic{footnote}}
\setcounter{footnote}{0}

The Minimal Supersymmetric Standard Model (MSSM) makes a firm
prediction for the mass of the lightest Higgs boson.  
Supersymmetry (SUSY) relates the quartic
coupling of the Higgs boson to the
Standard Model (SM) gauge couplings, resulting in a tree-level
prediction for the Higgs mass: $m_{h} < M_{Z} = 91 $ GeV.
The current Higgs mass bound from LEP II
($m_{h}\ge 114.4 \text{ GeV}$ \cite{Barate:2003sz})
can be accommodated, but only  if the parameters are somewhat 
fine-tuned --- proponents
of the MSSM would have been more comfortable had the
Higgs boson been discovered closer to the tree-level prediction.
(See \cite{Kane:2004tk} for a discussion of fine-tuning in the MSSM.)
The alternative to fine-tuning is 
new physics at the TeV scale that contributes to the Higgs mass.  
For recent attempts, see 
\cite{DEKPuneet,DEKPuneet2,Harnik:2003rs,Birkedal:2004xi,Polonsky}.

There are two hints as to the nature of this new TeV scale physics.  The
first is the striking unification of the gauge couplings in the
MSSM \cite{Savas}.   
We will demand that any modifications to the Higgs sector
maintain unification.
Second, the theory be should as natural as possible;
some mechanisms raise the Higgs boson mass only by
introducing a substantial fine-tuning.  
As we will discuss in section 2.1, these two
criteria lead us to study particular gauge extensions of the MSSM.
In these extensions the Higgs mass is increased through 
non-decoupling $D$-terms \cite{DEKPuneet,LR}.  
To have a significant effect, the new gauge group under which 
the Higgs is charged must have a large coupling.

Unfortunately, the simplest gauge extensions of the MSSM spoil
gauge coupling unification.  In these cases, the only recourse
is to include additional particles whose sole purpose to restore
unification, ``unifons.''   In this paper, 
using the success of the MSSM as a guide, we will describe two models
without these designer particles.
In these models, gauge coupling unification constrains the size of the 
non-decoupling $D$-terms, limiting the potential increase in the Higgs mass.  
This is easy to see: in our approach, we mix
the electroweak $SU(2)\times U(1)$ with additional gauge groups.  This  
increases the $SU(2)\times U(1)$ gauge couplings, which are
related to the Higgs quartic coupling by supersymmetry.
Unification relates these new electroweak gauge groups 
to a new colored gauge group, which is in danger of becoming strongly
coupled.
This puts an upper bound on the size of the 
$SU(2)\times U(1)$ coupling,
and therefore on the $D$-term contribution to the Higgs mass.

In Sec. 2 we present two gauge extensions of the MSSM that naturally maintain 
unification, and can contribute to the Higgs mass without invoking fine tuning.
The first model adds an extra copy of a GUT gauge group, which is coupled
to the standard model gauge content by bi-fundamentals.  We call this 
approach ``product unification.''  The second model is one of
accelerated unification \cite{Arkani-Hamed:2001vr}, 
where the additional gauge content is a copy of the standard model gauge group.  
In this case unification is maintained, but occurs at a much lower scale.
Notably, this model requires the presence of a second pair of Higgs doublets 
at low energy, which might seem an {\it ad hoc} addition
to the model, resurrecting the ``unifon'' specter.  On the contrary, 
we will argue in Sec. \ref{Sec: GUT Scale} that their 
existence can be related to the observed
stability of the proton via a missing-partner mechanism 
\cite{MissingPartner}.

The organization of the rest of the paper is as follows.  In Sec. 3
we discuss in detail non-decoupling $D$-terms, focusing 
on the specific potential that arises in both of the models described
above.
In Sec. 4 
we apply these results to the product unification model 
and discuss implications for the Higgs mass.  We find that in this case
the non-decoupling $D$-terms can contribute roughly 10 $\GeV$ to the Higgs
mass.   In Sec. 5 
we turn to the minimal accelerated unification model, and conclude that 
$D$-terms can easily raise the Higgs mass by $30-40$ $\GeV$.
We discuss precision unification in both cases. 
Finally we conclude and present an outlook for these two models.

\section{Motivation: The Higgs Mass in Unified Models}

In Sec. \ref{Sec: Higgs Mass}, we will begin by outlining 
various modifications to the MSSM that can raise the Higgs mass.  
Of these, we  will argue that only two -- the NMSSM \cite{NMSSM} and 
non-decoupling $D$-terms -- can accommodate unification without fine tuning.
We will focus on the second of these, and give 
in Sec. \ref{Sec: Unification} a more
detailed description of the restrictions that unification places on
this approach.
There are two distinct possibilities: 
the unification scale may be preserved by the new gauge structure, or it may be
lowered.  In Sec. \ref{Sec:  Models} we briefly describe the two simplest 
implementations of these alternatives.  The first involves adding an extra
unified gauge group, while the second requires the addition of a
copy of the standard model gauge content.
 
\subsection{Increasing the Higgs Mass}
\label{Sec: Higgs Mass}

We will now discuss various ways of raising the Higgs mass in supersymmetric
models, and conclude 
that gauge extensions and the NMSSM are the two most attractive
alternatives.

\subsubsection*{SUSY Breaking in the MSSM}

The simplest way to increase the Higgs 
mass requires no new physics; it simply uses the SUSY breaking effects 
associated with the top squark \cite{MSSMHiggsBound}.
In general, a Yukawa interaction between the Higgs and some other particle
will contribute both to the (mass)$^{2}$ of the Higgs boson and to its 
quartic coupling, as
\begin{eqnarray}
\label{eqn:FT}
\delta \lambda = \frac{ N_c y^4}{8\pi^2} \left(
\log \frac{m_{\text{B}}}{m_{\text{F}}} + a^2\left(1 - \frac{a^2}{12}\right)\right),
\hspace{0.5in}
\delta m^2 = -  \frac{N_c (1+ a^2) y^2}{8 \pi^2} \left( m^2_{\text{B}} - 
m^2_{\text{F}}\right)
\log \Lambda^2
.\end{eqnarray}
Here $N_c$ is the number of colors, $m_B$ and $m_F$ are the boson and 
fermion masses, and we have included the SUSY violating $A$-term
\begin{eqnarray}
\LL_{\text{soft}} \supset y\,a\, m_B\; h \tilde{f} \tilde{f}^c.
\end{eqnarray}
At loop level, these effects give a logarithmically divergent
contribution to the Higgs boson (mass)$^{2}$,
but only a finite contribution to the Higgs quartic coupling.  The result
is fine-tuning, making this mechanism rather unattractive.  
For example, every 10 GeV increase in the Higgs mass
above 115 GeV requires a doubling of the top squark mass.

We could use the $A$-term to improve this situation, but such
contributions are typically small.  The $A$-term contribution
is maximized at $a = \sqrt{6}$, 
a parameter choice that has been dubbed ``maximal mixing.''
SUSY breaking scenarios often have much smaller $A$-terms,
reducing potential contributions to the Higgs mass\footnote{ 
For example, anomaly mediation and gravity mediation lead to
$a \sim \OO(1)$, gaugino mediation to $a\sim \OO(\alpha^\half)$, and
gauge mediation to $a\sim \OO(\alpha)$.  Dilaton mediation gives the largest 
value, $a \sim \sqrt{3}$ \cite{Kane:2004tk}.}.   
For example, as $a$ decreases from $\sqrt{6}$ to $1$ the contribution to $\delta 
\lambda$
decreases by a factor of 3. 
Moreover, radiative corrections from $A$-terms increase fine tuning in
the MSSM.  
We therefore conclude that $A$-terms are not useful for raising the Higgs mass.

\subsubsection*{Additional Matter}

It is possible to add Yukawa interactions to new particles, but
precision electroweak constraints make this unlikely.   
Chiral multiplets that get their masses from electroweak symmetry breaking 
are very strongly constrained.  Vector-like matter makes the already 
modest SUSY-breaking logarithm in Eq.~(\ref{eqn:FT}) even smaller, so is not 
a useful alternative. 

\subsubsection*{New $F$-Terms}

A more attractive way of raising the Higgs mass is to add new fields 
that directly couple to $h_u$ or $h_d$ in the superpotential.
At the renormalizable level the 
only possibilities are a gauge singlet, $n$, and an SU(2) triplet, $T$.  
The possible superpotential terms are
\begin{eqnarray}
\label{eqn:superpot}
 W = \left(
 \begin{array}{cc} h_u & h_d\end{array}\right)
 \left(\begin{array}{cc} \kappa_+ T_+ & \kappa_n n  + \kappa_0 T_0\\
 \kappa_n n + \kappa_0 T_0 & \kappa_- T_-\end{array}\right)
 \left( \begin{array}{c} h_u \\h_d\end{array}\right),
 \end{eqnarray}
where we have included three different triplets $T_\pm, T_0$ 
with hypercharges $\pm 1$, $0$.
These interactions contribute to the quartic coupling of the 
Higgs bosons at the tree-level, so can raise the Higgs mass without
fine-tuning \cite{Espinosa}.  
However, the triplets will 
disturb gauge coupling unification unless additional matter is added to 
fill out an SU(5) multiplet.  
The smallest such multiplets are the {\bf 24} 
of SU(5) for $T_{0}$, and the {\bf 15+}$ \mathbf{\overline{15}}$ 
for $T_{\pm}$.  
Absent an obvious rationale for adding 
the remainder of the GUT multiplets, we find this possibility somewhat 
distasteful.  Moreover, the addition of so much matter raises the prospect 
of a Landau pole.  Finally, 
if the triplets acquire a small vacuum expectation value (vev), they can have 
dangerous contributions to the $T$ parameter. On the other hand, 
the presence of a singlet $n$ will not affect the unification 
of couplings at the one-loop level, making it a more promising candidate.  
This coupling has a positive beta function and will lead to a Landau pole
at large coupling.  
This constrains the size of the $n h_{u} h_{d}$ coupling at the weak scale, 
thus limiting its contribution to the Higgs mass \cite{KaneKolda}.

\subsubsection*{New $D$-Terms}

Finally, one can introduce new gauge groups.
The associated $D$-terms will then contribute to the Higgs quartic coupling.  
In the SUSY limit,
these new contributions exactly decouple when the gauge groups break
down to the MSSM.  On the other hand, if the scale of SUSY breaking is 
close to the scale of the breaking of the gauge groups, 
there are non-decoupling $D$-terms which can raise the Higgs 
mass  -- essentially, when integrated out 
these terms introduce hard SUSY breaking into the
MSSM  \cite{DEKPuneet}.   

These contributions to the quartic 
coupling (and hence the physical Higgs boson mass) arise at 
tree level, while contributions to the (mass)$^{2}$ occur only at loop level.  
This allows an increase in the Higgs boson mass 
without significant fine-tuning.  
However, this new 
non-supersymmetric quartic coupling generates 
a quadratic divergence in the Higgs mass, which is cut 
off at the mass of the new vector bosons.  
To prevent fine tuning, we therefore require that this contribution not be
too large, since
\begin{eqnarray}
\frac{\delta \lambda}{\lambda + \delta \lambda} \frac{ M_V^2}{16\pi^2 v^2}\approx 
\text{Fine Tuning}^{-1} .
\end{eqnarray}
This constrains the 
breaking scale to be in the 3 -- 10 TeV range. The lower limit
is set by precision electroweak considerations.

We have seen that non-decoupling $D$-Terms have the potential to raise the
Higgs mass without fine-tuning, but we must still require 
that the new fields do not upset unification.
We will now proceed to outline the two different ways that extended 
gauge sectors can accomplish this, and present two minimal 
implementations of this mechanism.

\subsection{Unification in the MSSM}
\label{Sec: Unification}

At one loop, gauge coupling unification can be tested by examining the 
following relation between the inverse gauge couplings and the one loop 
beta functions for the gauge groups:
\begin{eqnarray}
\label{eqn:testfn}
\frac{\alpha^{-1}_3(M_{Z^{0}}) -\alpha^{-1}_2(M_{Z^{0}})}
{\alpha^{-1}_2(M_{Z^{0}})-\alpha^{-1}_1(M_{Z^{0}})} 
= \frac{b_3 - b_2}{b_2-b_1}\equiv B^{32}_{21},
\end{eqnarray}
where $b_i$ are the one loop beta function
coefficients.  Using the experimentally measured values of the gauge 
couplings at the weak scale \cite{PDG}, 
\begin{equation}
\label{Eqn:Measured}
\begin{array}{lll}
\alpha_{EM}^{-1}(M_{Z^0}) & =& 127.918 \pm 0.018 \\
\sin^{2} \theta_{W} (M_{Z^0}) &=& 0.23120 \pm 0.00015 \\
\alpha_{s}(M_{Z^0})  &=& 0.1187 \pm  0.0020
\end{array}
\end{equation}
the left hand side of Eq.~(\ref{eqn:testfn}) is $0.719 \pm 0.004$\footnote{
We have converted here $\alpha_{Y}$ to the GUT 
normalized $\alpha_{1} =(5/3) \alpha_{Y}$.}. 
In the MSSM, we have $(b_{1}, b_{2}, b_{3})=(33/5,1,-3)$, so
$B^{32}_{21}=0.714$.  This agreement summarizes the success of 
gauge coupling unification in the MSSM at one-loop.

We must now ask what mechanisms allow us to raise the Higgs
mass without changing $B^{32}_{21}$.  
One possibility is to add extra matter only in complete 
GUT multiplets.  Then all the $b_i$ are all shifted by a fixed
amount -- in this case the unification scale is unchanged, but the
value of the gauge couplings at unification may be altered.
The NMSSM is a trivial implementation of this strategy 
(all $b_{i}$ are unchanged),
and, as described above, is effective in raising the Higgs mass.
If we wish to consider gauge extensions to the MSSM, 
we must add a complete unified gauge group to the model.
A model of this form, which we refer to as product unification,  
will be described in the next subsection,
and in greater detail in Sec.~\ref{Sec: Product}.

The other natural possibility that keeps $B^{23}_{12}$ unchanged is to 
add matter in such a way that the $b_3-b_2$ is changed, but a
proportional change is made in $b_2-b_1$.  
In this case unification still occurs, but at a lower scale.  A specific 
implementation of this idea is accelerated unification
\cite{Arkani-Hamed:2001vr}, where
the unification scale is brought down to the intermediate scale.  The 
extra gauge content is simply another copy of 
$SU(3)\times SU(2)\times U(1)$.  A model 
of this form will be described below, and in more detail in 
Sec.~\ref{Sec: AT}.

\subsection{Two Minimal Gauge Extensions of the MSSM}
\label{Sec: Models}

We will present two minimal models that contain non-decoupling 
$D$-terms and preserve unification.  Product unification adds a full 
GUT gauge group; accelerated unification adds a second copy of the MSSM
gauge group.

Both of these are closely related
to deconstructed dimensions \cite{Deconstruction}.  
The first model, product unification,
is equivalent to having an extra dimensional GUT 
with $SU(3)\times SU(2)\times U(1)$
gauged on the boundary.    The second model is equivalent to a bulk 
$SU(3)\times SU(2)\times U(1)$
gauge theory with  power-law ``unification'' at a low scale.
In order to have non-decoupling $D$-terms the effective radius 
must be 1 -- 10 TeV, 
meaning that the bulk of the running 
occurs above the naive five dimensional cut-off. 
In both examples we will consider the minimally deconstructed theories, which
we now describe.

\subsubsection{Product Unification}
\label{Sec: P Model}

\begin{figure}
\begin{center}
\epsfig{file=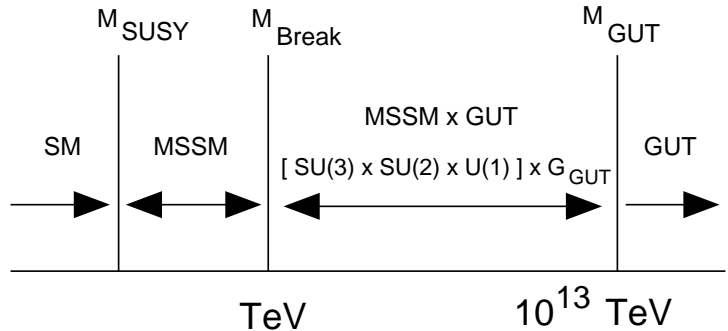,height=1.75in}
\caption{
\label{fig:diagonalP}
A minimal model of product unification.
The extended gauge group breaks down
to the diagonal subgroup (the SM gauge group) at the TeV scale. }
\end{center}
\end{figure}

In this approach, the high energy gauge group is 
$G = SU(3) \times SU(2) \times U(1)$, augmented by a grand 
unified gauge group, $G_{GUT}$.  
Near the TeV scale, the product $G \times G_{GUT}$ is broken down to the 
standard model gauge group, $G_{SM}$ (See Fig.~\ref{fig:diagonalP}).  
The matter and Higgs fields 
of the standard model are charged under $G$.  After the breaking, they inherit
the usual standard model quantum numbers.  

\begin{figure}
\begin{center}
\epsfig{file=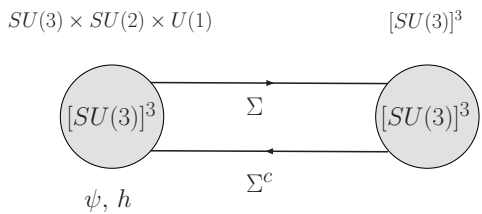,width=2.5in}
\caption{
\label{fig:Pmodel}
A theory space diagram of the product unification model discussed in the text. 
In addition to the usual $SU(3) \times SU(2) \times U(1)$, there is an
$SU(3)^3$ gauge group. 
The $\Sigma, \Sigma^{c}$ fields are bi-fundamentals connecting these
groups.  The three generations of matter, denoted $\psi$, 
and the pair of Higgs doublets, $h$, are charged under 
$SU(3)\times SU(2)\times U(1)$.  We consider a model where
$SU(3) \times SU(2) \times U(1)$ unifies
into the trinified group $SU(3)^3$ at the GUT scale. }
\end{center}
\end{figure}

The breaking to the MSSM occurs when link fields, $\Sigma$ 
and $\Sigma^c$, acquire a vev.  These fields transform 
as bi-fundamentals of the global symmetry associated with the GUT 
gauge group.    The structure of this model 
is similar to that of minimal deconstructed gaugino 
mediation \cite{DeconGauge};  however, we will 
remain agnostic about the exact mechanism of supersymmetry 
breaking\footnote{
Gaugino mediated SUSY breaking, along with a TeV diagonal breaking
scale, would give a too light $\tilde{\tau}$, unless we take the 
gauginos masses to be unnaturally heavy \cite{Transparent}.}.

While any GUT representation for the link fields will leave unification 
undisturbed, here we take the fields to transform under a 
trinified \cite{Trinified} representation (See Fig.~\ref{fig:Pmodel}).  
The reasons are two--fold.  First, this representation is the 
smallest possible.  
In trinification, the $\Sigma$ fields fall into representations 
of $SU(3)$ that only contribute $\Delta b = 3$  to each beta function.   
In  $SU(5)$ and  $SO(10)$ unification, 
the link fields add 5 and 10 to $\Delta b$,
respectively. Thus trinification contributes the least possible 
amount to the gauge coupling beta functions, which helps keep the theory 
perturbative. 
Second, this model is closely related to the 
minimal accelerated unification model, which we now discuss. 

\subsubsection{Accelerated Unification}
\label{Sec: MAT}
\begin{figure}
\begin{center}
\epsfig{file=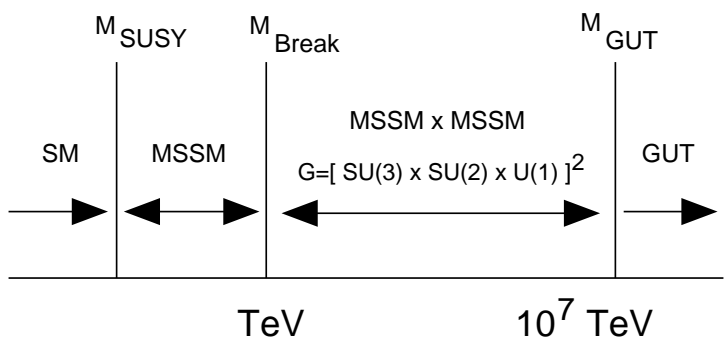,height=1.75in}
\caption{
\label{fig:diagonalA}
The minimal accelerated unification model, with $\NN=2$,
where two copies of the MSSM gauge group break down
to the diagonal subgroup at the TeV scale. }
\end{center}
\end{figure}

In accelerated unification models \cite{Arkani-Hamed:2001vr},
the Standard Model gauge group, $G_{\text{SM}}$, is the remnant of 
an enlarged group, $G_{\text{SM}}^{\NN}$, that 
breaks to the diagonal subgroup at the TeV scale.
The presence of extra matter changes the gauge coupling beta
functions, causing the theories to unify at a much lower scale  
(see Fig.~\ref{fig:diagonalA}).

\begin{figure}
\begin{center}
\epsfig{file=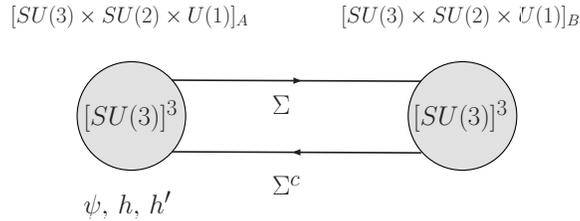,width=3in}
\caption{
\label{fig:model}
A theory space diagram of Minimal Accelerated Trinification. 
There are two copies of $SU(3) \times SU(2) \times U(1)$, 
labeled by subscripts $A$ and $B$, each of which
unifies into $SU(3)^3$.
The $\Sigma, \Sigma^{c}$ fields are bi-fundamentals connecting these
groups.  The three generations of matter, denoted $\psi$, 
and two pairs of Higgs doublets ($h$, $h'$) are also shown.}
\end{center}
\end{figure}

The gauge and matter content of the $\NN=2$ trinified
model is summarized in Fig.~\ref{fig:model}. 
There are two copies of the low energy gauge group, which we denote 
$[SU(3)_C \times SU(2)_{L} \times U(1)_{Y}]_{A,B}$.  
The matter and Higgs bosons of the MSSM, as well as a new pair of Higgs 
bosons, are charged 
under $[SU(3)_C \times SU(2)_{L} \times U(1)_{Y}]_{A}$.  Again, a vector-like 
pair of link fields, $\Sigma$ and $\Sigma^c$,
is responsible for breaking the gauge groups down to the diagonal
subgroup.  These fields should form complete GUT multiplets, 
so as to not contribute to the relative running of the gauge 
couplings.  A similar situation occurs in 
theories of gauge mediated supersymmetry breaking, where 
additional complete multiplets
are used so as not to spoil unification.  Because these fields make up 
full GUT multiplets, rather than the minimum necessary for the breaking of the
gauge symmetry, some components of the $\Sigma$ fields 
can become pseudo-Goldstone bosons (PGBs).  These particles can have important
phenomenological consequences, which we will address in Sec.~5.

While accelerated unification can accommodate any GUT representation
for the link fields, trinification is particularly elegant.
$SU(5)$ and $SO(10)$ models contain gauge boson mediated
dimension-six proton decay operators, now dangerous due to the lower 
GUT scale, which are difficult to remove in 4-$D$ GUT models.  
Moreover, the dynamics of the breaking 
$G_{SM}^2 \to G_{SM}$ is simplifier in trinified models, since
it is possible to stabilize the potential for $\Sigma$ in the $D$-flat
directions by adding renormalizable terms to the superpotential.
This is in contrast with the $SU(5)$ case, where there is only 
a $D$-term potential; no stabilizing superpotential can be 
added without additional matter \cite{Arkani-Hamed:2001vr}. Directions
that are not $D$-flat would lead to fine-tuning.
In addition, as already described above, trinification has the 
smallest representation for the $\Sigma$ fields, ameliorating Landau 
pole issues.  

We must also choose $\NN$, the
number of copies of the MSSM gauge group.
In principle we may add as many copies as desired, as long as $2 \NN$ 
Higgs doublets are added at the same time.  
But the expected threshold corrections to unification 
grow with $\NN$, so at large $\NN$ unification appears accidental.  
Moreover, the 
gauge couplings scale as $\sqrt{\NN} g_{SM}$, so strong coupling problems
can arise at large $\NN$.  For these reasons we focus on $\NN=2$, where we 
expect to have the best control over 
unification\footnote{For $\NN=3$, there is the interesting  
possibility that the three pairs of Higgs doublets are 
related to the three families (i.e. Higgs-Matter unification), but we will 
not explore this model here.}.

Finally, the $\NN=2$ trinified model presented here has an elegant
mechanism to prevent proton decay.  As noted above, this model includes
an extra pair of Higgs doublets in addition to the usual pair present in
supersymmetric models.  We may then couple one pair of Higgs doublets to
the leptons and the other to the quarks, which suppresses baryon
number violating interactions. 
We will discuss this mechanism in more detail in Sec.~\ref{Sec: P Decay}.

\section{Non-Decoupling D-Terms}
\label{Sec: DTerms}

\label{Sec: P TeV Scale Breaking}
In this section we discuss the breaking of
the extended gauge sector down to the SM gauge group.
The breaking is essentially identical for the product and accelerated 
unification models, and will ultimately be the source of the 
non-decoupling $D$-Terms that raise the Higgs mass.

Breaking occurs when the link fields $\Sigma, \Sigma^c$ get a vev.
In both of the models described above, 
the link fields can be organized into global $[SU(3)^3]_{A,B}$ multiplets as
\begin{eqnarray}
\nn
&&\Sigma_{C } \sim (\mathbf{3}_{CA},\overline{\mathbf{3}}_{CB})
\hspace{0.4in}
\Sigma_{C}^c \sim (\mathbf{3}_{CB},\overline{\mathbf{3}}_{CA})
\\
\nn
&&\Sigma_{L } \sim (\mathbf{3}_{LA},\overline{\mathbf{3}}_{LB})
\hspace{0.44in}
\Sigma_{L}^c \sim (\mathbf{3}_{LB},\overline{\mathbf{3}}_{LA})
\\
&&\Sigma_{R} \sim (\mathbf{3}_{RA},\overline{\mathbf{3}}_{RB})
\hspace{0.41in}
\Sigma_{R}^c \sim (\mathbf{3}_{RB},\overline{\mathbf{3}}_{RA}).
\end{eqnarray}
The hypercharge generator is given by
\begin{eqnarray}
Y = -\frac{1}{6} \,T^8_L  -  \frac{1}{3}\, \tilde{T}^8_R,
\end{eqnarray}
with $T^8_{L} =\diag(1,1,-2)$ and $\tilde{T}^8_R=\diag(-2,1,1)$.
These fields come in a complete GUT multiplet and do not disturb 
unification. 

\subsection{Decoupling the $D$-Terms}

To give the $\Sigma$ fields a vev, we must include a potential $V(\Sigma)$.
The most general $SU(3)$ symmetric superpotential is
\begin{eqnarray}
\label{Eq: SymSuperpotential}
W_{\Sigma} =   \lambda (\det \Sigma + \det \Sigma^c) + \mu \Tr \Sigma \Sigma^c.
\end{eqnarray}
There is one such potential for each of ${\Sigma_C,\Sigma_L,\Sigma_R}$.  
The Kahler term is
\begin{eqnarray}
K = \Tr e^{g_A V_A} \Sigma e^{-g_B V_B} \Sigma^\dagger
+ \Tr e^{g_B V_B} \Sigma^c e^{-g_A V_A} \Sigma^c{}^\dagger .
\end{eqnarray}
This potential has two  $D$-flat minima at:
\begin{eqnarray}
\label{Eq: VEVs}
\langle\Sigma\rangle=\langle\Sigma^c\rangle =  0 \hspace{0.4in}
\langle\Sigma\rangle=\langle\Sigma^c\rangle =  - \frac{\mu}{\lambda} 
\identity\equiv f \identity.
\end{eqnarray}
We focus on the second solution, which breaks 
$SU(3) \times SU(3) \rightarrow SU(3)$.  The small 
fluctuations around the vev can be 
grouped into four complex fields $S$, $\eta$, $\pi$ and $\phi$:
\begin{eqnarray}
\label{Eq: Decomposition}
\nonumber
\Sigma & =& e^{\eta/\sqrt{6}f}\exp(c_A\pi /f) \Sigma_0 \exp(c_B \pi/f) \\
\Sigma^c &=& e^{-\eta/\sqrt{6}f}\exp(-c_B \pi/ f) \Sigma_0 \exp(-c_A \pi/ f) \\
\nonumber \Sigma_0 &\equiv&  \Big(f + \frac{S}{\sqrt{6}} \Big) \identity + \phi, 
\end{eqnarray}
where $c_{A,B}$ are normalization constants. 
$S$ is the fluctuation of the vev, while $\eta$ and $\pi$  are 
the Goldstone boson
superfields for the broken global $U(3)$ symmetries.  Only a 
subgroup of this $U(3)$ is gauged.  The determinant 
superpotential breaks the $U(1)$ of this $U(3)$ explicitly, giving 
$\eta$ a mass.   The fields transform under the unbroken $SU(3)$ as
\begin{eqnarray}
S \sim \mathbf{1}, \hspace{0.3in}
\phi \sim \mathbf{8}, \hspace{0.3in}
\eta \sim \mathbf{1}, \hspace{0.3in}
\pi \sim \mathbf{8}. 
\end{eqnarray}
Expanding the superpotential around the vev,  we find: 
\begin{eqnarray}
W =  2 \lambda \cosh \sqrt{\frac{3}{2}}  \frac{\eta}{f}  \det \Sigma_0 + \mu \Tr 
\Sigma_0^2 .
\end{eqnarray}
This yields masses
\begin{eqnarray}
\mu_S = -\mu, \hspace{0.3in}
\mu_\phi = 2\mu, \hspace{0.3in}
\mu_\eta = -3\mu , \hspace{0.3in}
\mu_\pi = 0. \hspace{0.3in} 
\end{eqnarray}
The $D$-terms decouple in the supersymmetric
limit.  To see this,  define the vector superfields $V_0$ and $V_H$ by
\begin{eqnarray}
g_A V_A = g_0 V_0 - g_0 \frac{g_A}{g_B} V_H\hspace{0.5in} 
g_B V_B = g_0 V_0 + g_0 \frac{g_B}{g_A} V_H,
\end{eqnarray}
with $g_0^{-2} = g_A^{-2}+g_B^{-2}$.  It follows that $V_H$ acquires a mass
while $V_0$ remains massless.  
Ignoring $V_0$, the Kahler term may be expanded to leading order as
\begin{eqnarray}
 K= 2 \Tr \phi^\dagger \phi + \Big|f + \frac{S}{\sqrt{6}}\Big|^2 (6  +  \Tr| g_H 
V_H+(\pi+\pi^\dagger)/f| ^2)  + \eta^\dagger \eta
  + \cdots
\end{eqnarray}
where $g_H^2= g_A^2 + g_B^2$. 
Using the gauge transformation $V \rightarrow V+ \alpha +\alpha^\dagger$ 
we can go to unitary gauge, with $\pi=0$.  Now consider a field, $H$, charged 
under $G_A$. The Kahler term contains
the coupling to $V_H$ (for the moment we suppress the $e^{g_0 V_{0}}$)
\begin{eqnarray}
K= H^\dagger e^{g_A V_A} H = H^\dagger H -\frac{g_0g_A}{g_B}  H^\dagger  V_H H+ 
\cdots
\end{eqnarray}
The superfield propagator
at zero momentum in unitary gauge is given simply by $1/M_V^2$. So, 
integrating out $V_H$  gives  
\begin{eqnarray}
K_\eff=H^\dagger H -  \frac{g_0^2 g_A^2}{g_B^2}\frac{1}{g_H^2 f^2}|H^\dagger T^a 
H|^2 + \cdots
\end{eqnarray}
The second term contains several interactions that  
provide important constraints on our models,
but no scalar potential.  
The scalar potential comes from restoring the $e^{g_0 V_{0}}$ to 
the first term. 
This is the just the standard decoupling
of the $D$-terms, automatic in the superspace formalism.  
In component field language, this decoupling
arises after integrating out the
$C$-component (i.e. the lowest component) of $V_H$, 
which corresponds to the lowest component of 
$\pi + \pi^\dagger$ in unitary gauge.

\subsection{Recoupling the $D$-Terms}

To avoid decoupling and increase the Higgs mass, we must include SUSY
breaking effects.  In unitary gauge, this is accomplished by giving 
the lowest component of $V_H$ a supersymmetry 
breaking mass. 
This can be done in the superspace formalism with a $D$-term 
spurion\footnote{
In principle, one could also add $\theta^{2}$ and ${\overline{\theta}}^{2}$ masses.}
\begin{eqnarray}
K= M_V^2(1 + \theta^4 m^2_{\text{soft}})  \Tr V_H^2 ,
\end{eqnarray}
with $M_{V}^{2} =g_{H}^{2} f^{2}$.  Integrating out $V_{H}$, we find the 
effective Kahler potential for $H$
\begin{eqnarray}
\label{eq:keff}
K_\eff = H^\dagger e^{g_0 V_0} H  - \frac{g_0^2 g_A^2}{g_B^2}\left( \frac{1}{g_H^2 
f^2}- 
\frac{  m^2_{\text{soft}} \theta^4}{g_H^2 f^2+ m^2_{\text{soft}}} 
\right)|H^\dagger T^a H|^2 + \cdots
\end{eqnarray}
where we have used the modified vector superfield propagator
\begin{eqnarray}
\Delta_F(p,\theta,\bar{\theta}) = -\frac{1}{p^2-M_V^2}  + 
\frac{m^2_{\text{soft}}\theta^4}{p^2-(M_V^2 + m^2_{\text{soft}})} + \cdots
\end{eqnarray}
In the limit $m^2_{\text{soft}}\rightarrow \infty$, 
the supersymmetry breaking coefficient in Eq.~(\ref{eq:keff}) is maximized,
and the Higgs quartic coupling becomes
(including the supersymmetric $D$-term from the 
unbroken gauge theory)
\begin{eqnarray}
\label{Eq: Non Decoupling D Term}
\lambda_{\text{Higgs}}=g_0^2\Big( 1+ \frac{g_A^2}{g_B^2}\Big)= g_A^2.
\end{eqnarray}
The quartic coupling is equal to the $D$-Term of the unbroken theory.

The $D$-term contributions the Higgs mass will be maximized if 
two conditions are satisfied.
First, SUSY breaking must be effectively 
communicated to the vector boson mass from the soft
Lagrangian.  Second, the gauge coupling $g_A$ must be large.    We 
will postpone the discussion of $g_{A}$ in specific 
models to Sec.~\ref{Sec: P Higgs Mass} 
and~\ref{Sec: A Higgs Mass}, where we will see that unification 
restricts its size.
Here we concentrate on how SUSY 
breaking is communicated to the vector boson soft mass.  First, we write 
down the most general soft Lagrangian for $\Sigma$ using spurions
\begin{eqnarray}
\label{Eq:spurion}
W&=&  (1+ a\theta^2) \lambda (\det \Sigma + \det \Sigma^c) + (1+ b\theta^2)\mu \Tr 
\Sigma\Sigma^c\\
K&=& (1+m^2 \theta^4) \left( \Tr e^{g_A V_A} \Sigma e^{-g_B V_B} \Sigma^\dagger
+ \Tr e^{g_B V_B} \Sigma^c e^{-g_AV_A} \Sigma^c{}^\dagger\right) .
\end{eqnarray}
Rewriting in terms of the physical fields, we have
\begin{eqnarray}
W&=& 2(1+ a \theta^2)\lambda \cosh \sqrt{\frac{3}{2}} \frac{\eta}{f}  \det \Sigma_0
+ (1+ b\theta^2) \mu \Tr \Sigma_0^2\\
K&=& (1+ m^2\theta^4)\left( 
 \Tr \phi^\dagger \phi +  \Big|f + \frac{S}{\sqrt{6}}\Big|^2 (6  +   g_H^2\Tr 
V_H^2)  + \half (\eta^\dagger+ \eta)^2
  + \cdots
\right) .
\end{eqnarray}
The lowest component of $V_H$ has already acquired a mass from $m^2$.
In principle, the addition of SUSY breaking can induce a tadpole for $S$.  
This can be removed by a $\theta$ dependent shift\footnote{
This $\theta$ dependent shift would induce a $\theta^2$ soft mass for the vector field, giving an effective $\theta^2 |H|^4$ interaction.}.  Such a shift can 
contribute to the vector mass (see Appendix B). For simplicity, we assume no tadpole is generated, 
which amounts to enforcing
\begin{eqnarray}
\label{Eq: NoTadpole}
m^2 +b\mu -a\mu =0 . 
\end{eqnarray}
Relaxing this condition can lead to additional sources of non-decoupling,
but this choice is sufficient to demonstrate that a significant 
non-decoupling is possible.
Taking $a=0$ in Eq.~(\ref{Eq: NoTadpole}), 
it follows that the masses of all the fields 
($S$, $\phi$, $\eta$) are positive for $m^2 > -\mu^{2}/2$. 
After integrating out the massive vector superfield the
effective action is
\begin{eqnarray}
K_\eff = H^\dagger e^{g_0 V_0} H +\frac{g_0^2g_A^2}{g_B^2} \frac{m^2 
\theta^4}{\left(\frac{g_H\mu}{\lambda}\right)^2 + m^2} (H^\dagger T^a H)^2 
+ \cdots 
\end{eqnarray}
In fact, as long as $\lambda$ is not too small, $m^2\sim \mu^2\sim - b\mu$ 
gives $\OO(1)$ re-coupling without destabilizing any modes.

\section{Product Unification}
\label{Sec: Product}

We now return to a more detailed discussion of the model of 
Sec.~\ref{Sec: P Model}.    
We begin with a discussion of one-loop running in this 
model, and derive the relations between low 
energy $SU(3)\times SU(2)\times U(1)$ parameters and 
the high energy $SU(3)^3$ parameters.
We then address the 
central question of the Higgs mass in \ref{Sec: P Higgs Mass}.
We close this section with a  discussion of unification beyond one-loop.

\subsection{One Loop Running}
\label{Sec: P Gauge Running}

This model unifies at one loop by construction.  
At one loop, the beta functions are
\begin{eqnarray}
\label{eqn:betafn}
\frac{d }{d t} \frac{8\pi^2}{g^2_i(t)} = -  b_{0,i}.
\end{eqnarray}
Here $i$ runs over the possible gauge groups, 
and the energy scale $t$ is defined to be 
$\log\left(\mu/ 3\text{ TeV} \right)$.  The coefficients $b_{0,i}$ are 
listed in  Table \ref{Tab: Beta P}.  The trinified gauge group 
$[SU(3)_{B}]^3$ starts at a unified coupling, and maintains unification under 
renormalization group (RG) flow.  We denote this coupling $g_{B}$.

Using the standard MSSM beta functions, we can run the measured gauge 
couplings in Eq.~(\ref{Eqn:Measured}) up to 3 TeV ($t=0$).
There we match on to the extended gauge sector via
\begin{eqnarray}
\frac{8\pi^2}{g^2_{i,\SM}(0)} = \frac{8\pi^2}{g^2_{i,A}(0)} + 
\frac{8\pi^2}{g^2_{B}(0)} = \frac{8\pi^2}{g^2_{i,A}(t)} + b_{i,A} t 
+ \frac{8\pi^2}{g^2_{B}(t)} +  b_{B} t.
\end{eqnarray}
Unification is maintained in this extension because the same 
quantity (the last two terms in the above equation) is
added to each of the SM gauge couplings.
In Sec. \ref{Sec: P Precision Unification} we will discuss higher order
corrections to unification.  Since the relative running of the gauge 
couplings is unaffected, the unification scale $t_{*}$ is unchanged:
$t_{*} =29.5$, corresponding to the usual $E_{GUT} \sim 10^{16} \GeV$.

\renewcommand{\baselinestretch}{1.25}
\begin{table}
\begin{center}
\begin{tabular}{|l||c|c|c||c|}
\hline
&$U(1)_A$& $SU(2)_A$& $SU(3)_A$&$[SU(3)_{B}]^{3}$\\
\hline
$V$&0&-6&-9&-9\\
$\psi$&6&6&6&0\\
$H$& $\frac{3}{5}$&1&0&0\\
$\Sigma$&3&3&3&3\\
\hline\hline
$b_0$&$9\frac{3}{5}$&4&0&-6\\
\hline
\end{tabular}
\caption{\label{Tab: Beta P} Beta function coefficients for the different
gauge groups.  }
\end{center}
\end{table}
\renewcommand{\baselinestretch}{1.0}

From  Table \ref{Tab: Beta P}, we see that the $SU(3)_B$ beta 
function coefficients are negative, so
the corresponding couplings run strong at low energies.  
The requirement that $SU(3)_B$ remain perturbative above the TeV scale
sets a maximum value for $g_{*,B}$, the value of the unified gauge
coupling at $t_*$.  If we define the low scale coupling
$8\pi^2/g^2_{B}(0)=\Delta$, then
\begin{eqnarray}
\frac{8\pi^2}{g^2_{*,B}} =\frac{8\pi^2}{g^2_{B}(0)} -  b_{B} t_*= 177 + \Delta.
\end{eqnarray}
We require $\Delta$ to be reasonably large to ensure that $SU(3)_{B}$ stays 
weakly coupled. 

\begin{figure}
\begin{center}
\epsfig{file=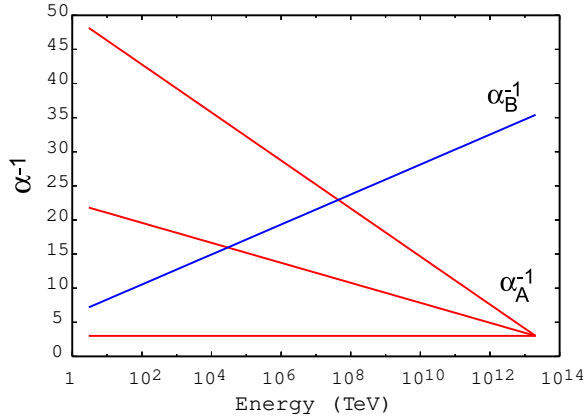,width=3in}
\caption{
\label{Fig: P Gauge Running}
The running of the gauge couplings, with $\alpha_{A}^{-1}=3$ at the GUT
scale.} 
\end{center}
\end{figure}

Similarly, $8\pi^2/g^2_{*,A}$ can be obtained by 
matching the $A,B$ gauge couplings onto the measured gauge couplings
at the weak scale,
\begin{eqnarray}
\frac{8\pi^2}{g^2_3(0)}= 64 =  \frac{8\pi^2}{g^2_{3,A}(0)}+ 
\frac{8\pi^2}{g^2_{B}(0)}=  
\frac{8\pi^2}{g^2_{*,A}}+ \frac{8\pi^2}{g^2_{*,B}} -  (b_{3,A}+ b_{B}) t_*  .
\end{eqnarray}
This implies that $8\pi^2/g^2_{*,A} = 64 - \Delta$.
In Fig.~\ref{Fig: P Gauge Running} we plot the one loop
running of the gauge couplings\footnote{
We could include GUT multiplets 
charged under the $[SU(3)]^{3}_{B}$ gauge group 
without spoiling unification.  However, once we fix the low energy
gauge couplings, these extra multiplets will contribute to low energy 
observables only at higher order, so may be neglected.
}.

\subsection{The Higgs Mass}
\label{Sec: P Higgs Mass}
From Eq.~(\ref{Eq: Non Decoupling D Term}) we see that 
the maximum fractional gain in the quartic depends on the gauge couplings. 
Making $g_{A}$ as large as possible, we find:
\begin{eqnarray}
\frac{\delta \lambda_{SU(2)}}{\lambda_{SU(2)}} 
=\frac{ g_{2,A}^2}{ g_{2,\SM}^2} - 1 \lsim \frac{1}{3} ,
\hspace{0.5in}
\frac{\delta \lambda_{U(1)}}{\lambda_{U(1)}} 
=\frac{ g_{1,A}^2}{ g_{1,\SM}^2} - 1 \lsim \frac{1}{7} .
\end{eqnarray}
We have used the values of the gauge couplings computed in the 
previous section -- higher loop corrections will be considered in
in Sec.~\ref{Sec: P Precision Unification}. 
The change in the quartic leads to a Higgs mass bound
\begin{eqnarray}
\label{eq:higgsbp}
m_{h^0}^2 \le \frac{(1.3 g^2 + 1.15 g'{}^2)}{g^2+ g'{}^2} M_{Z^0}^2
=(102 \GeV)^2,
\end{eqnarray}
a modest $11 \GeV$ gain over the MSSM tree-level prediction.
With top squarks of \mbox{400 GeV} the Higgs mass can be lifted to the 
LEP II bound of 114 GeV.  The fine-tuning of the MSSM may be ameliorated, but 
the Higgs cannot be made significantly heavier.

In this model $D$-term contributions to the Higgs mass
are tightly constrained.  This is because
the value of $g_{2,A}(M_{Z^0})$ 
is bounded by the total amount of relative running between
$SU(3)_C$ and $SU(2)_L$.   This conclusion is fairly robust in any model
with a product unification structure.  Adding either more gauge groups 
or matter charged under $[SU(3)]^3_B$ will not effect the relative running
between $SU(2)_{A}$ and $SU(3)_{A}$. 
One could charge some of the  
Standard Model matter (such as the first two generations)
under $B$ gauge groups rather than $A$, but this would cause
$SU(3)_{C,A}$ to run asymptotically free, tightening the bounds
on $g_{A}(\Lambda)$.  
The only way to increase the Higgs mass bound, Eq.~(\ref{eq:higgsbp}), 
is to charge the Higgs doublets under $[SU(3)]^3_B$.  In this
case, renormalizable Yukawa couplings are not possible, and some {\it ad hoc} 
change must be made to the $A$ sector to recover unification.

\subsection{Precision Unification}
\label{Sec: P Precision Unification}

While unification in this model is guaranteed at the one-loop level,
several effects may alter the accuracy of this prediction, such as 
higher loop contributions, 
TeV scale supersymmetric threshold corrections, GUT
scale threshold corrections, and SUSY breaking threshold corrections.
The second two are model dependent, depending in detail upon the
GUT scale physics, as well as the mechanism of supersymmetry breaking.
The first two are, however, calculable.  We now quantify the  
deviation from the one-loop prediction due to these effects.  

Fortunately, the holomorphicity of gauge couplings in supersymmetric
theories simplifies the analysis\footnote{
We stress that the method used here is equivalent to the more 
traditional approach of multi-loop running, where gauge 
couplings are matched at each mass scale.  
This formalism packages the results in an elegant way, but
is only applicable when SUSY breaking effects are small.}.
At first, it might appear that unification is disturbed by 
splittings of the $\Sigma$ masses, which are induced by renormalization
group evolution.  This could lead to a TeV scale supersymmetric threshold 
correction.  However, this correction cancels against certain two-loop
contributions.  The result is that higher loop effects can be encapsulated 
in the change in the anomalous dimensions of the light fields 
\cite{NSVZ,NimaHitoshi}.  

Integrating the NSVZ exact beta function, we can derive 
an RG invariant matching equation (see Appendix A).  We match the 
diagonal MSSM gauge coupling at the cutoff, $\Lambda$, and run down to 
low energies via
\begin{eqnarray}
\nonumber
\frac{8\pi^2}{g^2(\mu)} + C_2 \log g^2(\mu) &=& 
\frac{8\pi^2}{g_A^2(\Lambda)} + C_2 \log g_A^2(\Lambda)
+\frac{8\pi^2}{g_B^2(\Lambda)} + C_2^{B} \log g_B^2(\Lambda)\\
&&\nonumber
- C_2^{B} \log \frac{\Lambda^2}{\langle\Sigma \Sigma^c\rangle}
+ b_{\phi} \log \frac{\Lambda}{m_\phi}\\
&&
  + b_{0,\text{light}} \log \frac{\Lambda}{\mu}- \sum_{a} T_2(a)  \log 
Z_a(\mu,\Lambda).
\label{Eqn: LowGauge P}
\end{eqnarray}
The sum is over the light fields of the MSSM, $C_{2}$ and 
$C_2^B$ are quadratic 
Casimirs, $b_{0,\text{light}}$ is the one-loop beta function of the 
light fields, and $b_{\phi}=3$ is the contribution of the $\phi$ fields to the 
beta function.

Our goal is to find the multi-loop analog of $B^{32}_{21}$, defined
in Eq.~(\ref{eqn:testfn}). 
We start by considering the gauge couplings of 
the $SU(3)$ and $SU(2)$ sectors: 
\begin{eqnarray}
\label{Eq: P 32 Diff}
\Delta^{32}\equiv \frac{8\pi^2}{g_3^2(\mu)} - \frac{8\pi^2}{g_2^2(\mu)} &=& 
\Delta b^{32}_{\text{MSSM}}\log \frac{\Lambda}{\mu} 
+ 3 \log \frac{g^2_A(\Lambda)}{g^2_3(\mu)} - 2 \log \frac{g^2_A(\Lambda)}{g^2_2(\mu)} + \delta 
z_{\psi}^{32} +  \delta z_{h}^{32},
\end{eqnarray}
where $\Delta b^{32}_{\text{MSSM}}=-4$ 
is the difference between the one loop beta functions.
The final terms come from the $\log Z$s of the light fields:
\begin{eqnarray}
\delta z_\psi^{32} = \half \log \prod_{f=1}^{3} \frac{ Z_{u^c_f} Z_{d^c_f}}{Z_{q_f} 
Z_{l_f}},
\hspace{0.4in}
\delta z_h^{32} = -\half \log  Z_{h_u} Z_{h_d}, 
\end{eqnarray}
all of which are evaluated from $\Lambda$ down to $\mu$.
Similarly, for the $SU(2)$ and $U(1)$ couplings
\begin{eqnarray}
\label{Eq: P 21 Diff}
\Delta^{21}\equiv \frac{8\pi^2}{g_2^2(\mu)} - \frac{8\pi^2}{g_1^2(\mu)} &=& 
\Delta b^{21}_{\text{MSSM}}\log \frac{\Lambda}{\mu} 
+ 2 \log \frac{g^2_A(\Lambda)}{g^2_2(\mu)}
+ \delta z_{\psi}^{21} +  \delta z_{h}^{21} 
\end{eqnarray}
where
$\Delta b^{21}_{\text{MSSM}}=-5\frac{3}{5}$.
The $\log Z$s of the light fields are given by
\begin{eqnarray}
\delta z_\psi^{21} = \frac{1}{5} \log \Pi_f \frac{ Z_{q_f}^7 Z_{l_f}}{Z_{u^c_f}^4 
Z_{d^c_f} Z_{e^c}^3},
\hspace{0.4in}
\delta z_h^{21} = \frac{1}{5} \log  Z_{h_u} Z_{h_d}.
\end{eqnarray}

We can now summarize the deviation from MSSM unification.  
In the MSSM\footnote{
In this case the numerical values of $\delta z$ change and
Eqs.~(\ref{Eq: P 32 Diff}) and (\ref{Eq: P 21 Diff}) are modified by 
the replacement $g_{A} \rightarrow g_{SM}$.}
\begin{eqnarray}
\frac{\Delta^{32}}{\Delta^{21}}&=& \frac{-4 -0.11 }{-\frac{28}{5}-0.05} \simeq
0.727.
\end{eqnarray}
Here we have evaluated the MSSM expression at moderate $\tan \beta$, where the
top Yukawa, $y_{t} \sim 1$, but other Yukawa couplings are insignificant.
This is to be compared to the experimental 
ratio of Eq.~(\ref{eqn:testfn}) (run up to 3 TeV and converted to the $\overline{DR}$ scheme) which yields $0.718$. However, corrections due 
to the SUSY breaking spectrum (in particular the 
mass splitting between the wino and gluino) can be significant.
In the product unification model, we find
\begin{eqnarray}
\frac{\Delta^{32}}{\Delta^{21}}
&=& \frac{-4 - 0.07}{-\frac{28}{5} -0.03} \simeq 0.722
\end{eqnarray}
for $\alpha_A^{-1} = 3$.  While both the numerator and 
denominator vary with $\alpha_A^{-1}$ and $\lambda$, the 
ratio is fairly insensitive to the choice of parameters.
We conclude that the calculable deviation from the 
MSSM prediction is 
roughly one $\sigma$. This is not particularly significant,
since the threshold effects from SUSY breaking and GUT scale 
physics are likely larger than the above deviation.  

\section{Accelerated Unification}
\label{Sec: AT}

In this section, we analyze the minimal model of accelerated unification
described in Sec.~\ref{Sec: MAT}.
In Sec.~\ref{Sec: A Gauge Running} we will discuss one-loop running.
We then describe the Higgs mass bounds in this model.  In 
Sec.~\ref{Sec: GUT Scale}, we discuss some basics of the GUT scale
physics, and conclude that the extra pair of Higgs
doublets can suppress proton decay in this model.  
Finally, we analyze unification beyond one-loop.

\subsection{One Loop Running}
\label{Sec: A Gauge Running}

Again, one loop unification is incorporated into this model by
construction.  The RGEs are given by 
Eq.~(\ref{eqn:betafn}), with coefficients $b_{0,i}$ listed in  
Table \ref{Tab: A Beta}.

As before, we use the MSSM beta function to run up to 3 TeV ($t=0$), where
we match on to the extended gauge sector via
\begin{eqnarray}
\frac{8\pi^2}{g^2_{i,\SM}(0)} = \frac{8\pi^2}{g^2_{i,A}(0)} + 
\frac{8\pi^2}{g^2_{i,B}(0)} = \frac{8\pi^2}{g^2_{i,A}(t)} + b_{i,A} t + 
\frac{8\pi^2}{g^2_{i,B}(t)} 
+  b_{i,B} t.
\end{eqnarray}
The unification scale is now given by
\begin{eqnarray}
t_* = \frac{\frac{8\pi^2}{g^2_{\SM,1}(0)} - \frac{8\pi^2}{g^2_{\SM,2}(0)}}{ 
\sum_i(b_{1,i} - b_{2,i})}= 15,
\end{eqnarray}
i.e. at energies  $E_{GUT} \sim 10^7 \TeV$.
The scale of unification has been lowered to the geometric mean of the 
MSSM GUT scale and the TeV  scale.

\renewcommand{\baselinestretch}{1.25}
\begin{table}
\begin{center}
\begin{tabular}{|l||c|c||c|c||c|c|}
\hline
&$U(1)_A$&$U(1)_{B}$&
$SU(2)_A$&$SU(2)_{B}$&
$SU(3)_A$&$SU(3)_{B}$\\
\hline
$V$&0&0&-6&-6&-9&-9\\
$\psi$&6&0&6&0&6&0\\
$H$& $\frac{6}{5}$&0&2&0&0&0\\
$\Sigma$&3&3&3&3&3&3\\
\hline\hline
$b_0$&$10\frac{1}{5}$&$3$&5&-3&0&-6\\
\hline
\end{tabular}
\caption{\label{Tab: A Beta} Beta function coefficients for the different
gauge groups.  }
\end{center}
\end{table}
\renewcommand{\baselinestretch}{1.0}

\begin{figure}
\begin{center}
\epsfig{file=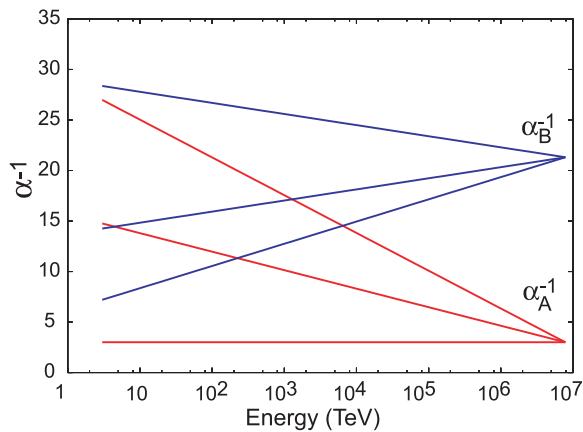,width=3in}
\caption{
\label{Fig: Gauge Running}
The running of the six gauge couplings, with $\alpha_{A}^{-1}=3$
at the GUT scale.
}
\end{center}
\end{figure}

Note that $SU(3)_{C,B}$ has a 
negative beta function coefficient, so runs strong at low energies.  
The requirement that $SU(3)_{C,B}$ remain weakly coupled above the TeV scale
sets a minimum value for $8\pi^2/g^2_{*,B}$, constraining the value of the 
unified gauge coupling at $t_*$.  Defining the low scale coupling
$8\pi^2/g^2_{3,B}(0)=\Delta$,
\begin{eqnarray}
\frac{8\pi^2}{g^2_{*,B}} =\frac{8\pi^2}{g^2_{3,B}(0)} -  b_{3,B} t_*= 90+\Delta.
\end{eqnarray}
Table \ref{Tab: IR GC} shows the various TeV scale gauge couplings as a 
function of $\Delta$.
Finally, in Fig.~\ref{Fig: Gauge Running} we plot the one loop
running of the gauge couplings.

\renewcommand{\baselinestretch}{1.3}
\begin{table}
\begin{center}
\begin{tabular}{|l||c|c|}
\hline
&$\frac{8\pi^2}{g^2_A(0)}$&$\frac{8\pi^2}{g^2_B(0)}$ \\
\hline\hline
$U(1)$&$206-\Delta $& $142+\Delta $\\
$SU(2)$&$123-\Delta$ &$ 59 +\Delta$\\
$SU(3)$&$64 -  \Delta$ &$\Delta$ \\
\hline
\end{tabular}
\caption{\label{Tab: IR GC} 
Gauge couplings of the different  gauge groups at 3 TeV,  where the full gauge 
group
$[SU(3) \times SU(2) \times U(1)]^{2}$
breaks to $SU(3)\times SU(2)\times U(1)$.}
\end{center}
\end{table}
\renewcommand{\baselinestretch}{1.0}

\subsection{The Higgs Mass}
\label{Sec: A Higgs Mass}

Before discussing the Higgs mass bound in this model, we should address
the basic structure of the Higgs sector in the presence 
of the extra doublets.
Electroweak symmetry breaking with four Higgs doublets 
is complicated, and a detailed discussion lies
beyond the scope of this work.   In \cite{Drees:1988fc} the supersymmetric 
four Higgs doublet model was studied in some detail.    

The Yukawa couplings for the Higgs are (ignoring neutrinos)
\begin{eqnarray}
W_{\text{Flavor}}= y_u q h_u u^c +y_d q h_d d^c + y_e l h_e e^c .
\end{eqnarray}
We need  $h_u$, $h_d$, and $h_e$ to acquire vevs,
which constrains the mass terms of the theory.
We start with a four Higgs doublet model, with superpotential
\begin{eqnarray}
\label{Eq:  mu}
W_{\mu} &=& \left(\begin{array}{cc}h_u &h_\nu\end{array}\right)
 \left( \begin{array}{cc}
 \mu_{ud}& \mu_{ue}\\\mu_{\nu d} &\mu_{\nu e}
   \end{array}\right)
\left(\begin{array}{c}h_d\\h_e\end{array}\right) .
 \end{eqnarray}
In the absence of off-diagonal terms, $\mu_{ue}$ and $\mu_{\nu d}$, the quark
and lepton Higgs sectors are completely isolated. 
Thus only $h_u$ and $h_d$  \begin{it}or\end{it} $h_e$ and 
$h_{\nu}$ acquire vevs.  We therefore need non-zero mixing terms.    

To get a limit on the Higgs mass, consider the case 
where one $\mu$ is much larger than the rest.  Then the corresponding 
pair of Higgs doublets may be integrated out, reducing the Higgs sector 
to the standard one with two Higgs doublets.    This
procedure does not lead fine tuning.
For the purposes of our discussion, we will give a large mass to
$h_d h_\nu$.
This decoupling limit also has the 
effect of suppressing the masses of the down type quarks relative to the 
up quarks.  Away from this limit many Higgs bosons become light, and 
the couplings to the gauge bosons are significantly altered.
In this case, the LEP limit for the Higgs mass may be altered over sizeable 
regions of the parameter space -- this question certainly warrants 
further investigation, but will not be pursued here.

For the remainder of this section, we assume the above decoupling limit 
applies.  The limit on the tree-level Higgs mass bound 
in minimal accelerated unification is significantly relaxed compared 
to the previous model:
\begin{eqnarray}
\frac{\delta \lambda_{SU(2)}}{\lambda_{SU(2)}} 
=\frac{ g_{2,A}^2}{ g_{2,\SM}^2} - 1 \lsim 1 ,
\hspace{0.5in}
\frac{\delta \lambda_{U(1)}}{\lambda_{U(1)}} 
=\frac{ g_{1,A}^2}{ g_{1,\SM}^2} - 1 \lsim 1 .
\end{eqnarray}
The new bound on the Higgs mass is
\begin{eqnarray}
m_{h^0}^2 \le 2M_{Z^0}^2=(128 \GeV)^2,
\end{eqnarray}
so we can easily accommodate the LEP II bound without fine-tuning.
This considerable improvement over the product unification model
is possible because both the $A$ and $B$ gauge couplings split in
accelerated unification.
This means that $SU(3)_B$ remains perturbative for much larger values
of the $SU(2)_A$ coupling, allowing a larger Higgs mass.

In fact, if we charge the second pair of Higgs doublets under the $B$ groups
rather than $A$ groups, the limit is relaxed even further.  However, as
we will see in Sec.~\ref{Sec: P Decay},
if both pairs of Higgs doublets are charged under the $A$ groups
the model has a natural mechanism to suppress proton decay.

We could also increase the Higgs mass bound by considering
$\NN\ge3$ models.  This reduces the 
relative running of the Higgs gauge groups, but lowers the GUT scale at the
same time.  In addition, as more gauge groups
are added, threshold corrections increase and precision
unification is lost.    In Sec. \ref{Sec: A Precision Unification} 
we will see that unification
is quite delicate in accelerated unification models, and 
becomes more so as $\NN$ is increased.  
Thus, there seems to be a tension between increasing 
the Higgs mass and ensuring accurate gauge coupling unification.

\subsection{GUT Scale Physics}
\label{Sec: GUT Scale}

At the unification scale the gauge groups unify into 
$(SU(3)_C \times SU(3)_L \times SU(3)_R)_{A,B}$.
The usual MSSM matter and Higgs fields combine with 
additional vector-like matter to form a chiral
$\mathbf{27}$ of $[SU(3)]^3$.  We will assume that this new exotic matter 
acquires a mass at the GUT scale, so is not relevant to our discussion.

At the unification scale the matter fields, $\Psi$, 
and Higgs fields, $\Phi$, 
form the chiral $\mathbf{27}$  of $SU(3)^3_A$
\begin{eqnarray}
\Psi_{Q}^\SM, \Phi_Q  \sim (\mathbf{3}_{C},\overline{\mathbf{3}}_{L})_A,
\hspace{0.4in}
\Psi_{L}^\SM, \Phi_L \sim (\mathbf{3}_{L},\overline{\mathbf{3}}_{R})_A,
\hspace{0.4in}
\Psi_{\Qc}^\SM, \Phi_\Qc \sim (\mathbf{3}_R,\overline{\mathbf{3}}_{C})_A.
\end{eqnarray}
The subscript indicates the transformation properties under the gauge charges.
Throughout this section 
capital Greek letters ($\Psi,\Phi,\Sigma$) will denote representations 
of the trinified
GUT group.  Lower case Latin letters ($q, h_u, h_d$) will denote 
fields that transform under $SU(3)\times SU(2)\times U(1)$.

\subsubsection{A Trinified NMSSM}
\label{Sec: Trin NMSSM}

The NMSSM is naturally embedded in trinification.  The Higgs multiplets
contain a singlet $\Phi_L$, often dubbed the neutretto. 
As discussed further in Sec.~\ref{Sec: P Decay}, the 
superpotential $\det \Phi_L$ gives rise to the NMSSM coupling $n h_u h_d$. 
The trilinear $n^3$ term does not typically appear. However, a source
term for the scalar can appear after SUSY breaking and cause $n$ to
acquire a weak scale vev \cite{Linear Term}.

We start with four Higgs bosons and  two singlets at the high scale,
and RG flow the superpotential
\begin{eqnarray}
\label{Eq:  NMSSM}
W_{\text{NMSSM}} &=& \left(\begin{array}{cc}h_u &h_\nu\end{array}\right)
 \left( \begin{array}{cc}(\kappa_q n_q + \half \kappa_q' n_l) &
  (\kappa_l' n_l +  \kappa_q' n_q)\\
    (\kappa_l' n_l +  \kappa_q' n_q)& (\kappa_l n_l + \half \kappa_l' n_q)
    \end{array}\right)
\left(\begin{array}{c}h_d\\h_e\end{array}\right)
 \end{eqnarray}
down to the low scale.
We add soft breaking $A$ terms, soft masses for all the fields, and  
linear soft terms for the singlets.    

The analysis of the previous section assumed that the sole new 
contribution to the quartic coupling
came from the gauge sector.  However, it is quite possible 
that an NMSSM-like structure might be a part of the trinified model 
presented here.  In this case, just as in the NMSSM, there is an additional 
contribution to the quartic coupling.
The size of this effect is constrained by the requirement that the coupling
not reach a Landau pole below the GUT scale\footnote{
This condition has recently been reexamined 
in \cite{Harnik:2003rs}.}. 
In this section we will describe how this bound is 
relaxed in models of  accelerated unification -- this occurs 
because the unification scale is lower.  

The one loop RGE for the NMSSM--like superpotential 
coupling $\kappa_n$  ($W \ni \kappa_{n} n \bar{h} h$) above the TeV scale is
\begin{eqnarray}
\label{eqn:kapparun}
\frac{d}{dt} \kappa_n \sim  \frac{\kappa_n}{16\pi^2}\left(4 \kappa_n^2 - 
g_{2,A}^2 - \frac{3}{5} g_{1,A}^2 \right).
\end{eqnarray}
We have neglected the contribution of SM Yukawa couplings, which depend on 
which $\kappa$ parameter is being studied.  This equation is readily solved 
in the limit $\kappa_n\gg g$, 
\begin{eqnarray}
\kappa_n^2(0) \lsim \frac{2\pi^2}{t_*} \Rightarrow \kappa_n\lsim 1.15. 
\end{eqnarray} 
This is a modest increase over the standard NMSSM coupling bound.
The lowered GUT scale has relaxed the bound on the quartic contribution.

A secondary effect is that the NMSSM coupling is supported
by running of the gauge interactions.  Accelerated unification
increases the gauge couplings,  which changes the bounds on 
$\kappa_n$ \cite{DEKPuneet2}.
However, it is not possible to get the maximal benefit described in 
\cite{DEKPuneet2} in the more restricted accelerated unification framework.  
From Eq.~(\ref{eqn:kapparun}), this would require increasing either the
$SU(2)$ or $U(1)$ gauge couplings.  However, the size 
of these couplings is restricted by the condition that 
$SU(3)_{C\;A,B}$ coupling, which is larger than
the $SU(2)$ and $U(1)$ couplings, must be small. 

To summarize, the NMSSM couplings can contribute to the Higgs quartic 
couplings in accelerated unification models. 

\subsubsection{Proton Decay and the Four Higgs Doublets }
\label{Sec: P Decay}

In trinified models there is a $\mathbb{Z}_3$ symmetry that relates the three
gauge couplings to each other.  This leads to the introduction
of proton decay.  This is a model dependent  feature -- for instance in 
some string inspired models there is no $\mathbb{Z}_3$ symmetry, 
and the unified gauge coupling is set by the vev of a dilaton.   Here, we 
will take the $\mathbb{Z}_3$ symmetry seriously, and consider implications 
for proton decay.  Proton decay occurs through the exchange of 
colored Higgs triplets (see \cite{Tuhin} for a recent study).  
These triplets will get a GUT-scale mass which, 
depending on the flavor structure of the model, may not be enough 
to suppress proton decay via dimension five and six operators, in which 
case further model building is necessary.  As we will see, the addition of a 
second pair of Higgs doublets can easily suppress proton decay in our 
accelerated unification model.

To see how this occurs, we must first discuss the implementation of flavor in 
this model.  There are two pairs of Higgs doublets in accelerated 
unification.  We will couple one pair, $\Phi_L^{l}$, to 
the leptons and the other pair, $\Phi_L^{q}$, to the 
quarks.  Assuming that the flavor structure 
obeys the $\mathbb{Z}_{3}$ symmetry, the MSSM Yukawa interactions 
are schematically given by the following superpotential terms
\begin{eqnarray}
\nonumber
W_{\text{flavor}}&=&
  y_Q\, (\Psi_Q^\SM \Phi^q_L \Psi_\Qc^\SM
  +\Psi_L^\SM \Phi^q_\Qc \Psi_Q^\SM+\Psi_\Qc^\SM \Phi^q_Q \Psi_L^\SM)
  \\&&
+ y_L\, (\Psi_L^\SM \Psi_L^\SM \Phi^l_L  +\Psi_Q^\SM \Psi_Q^\SM \Phi^l_Q
+\Psi_\Qc^\SM \Psi_\Qc^\SM \Phi^l_\Qc) .
\end{eqnarray}
At the level of the superpotential, baryon number symmetry is exact 
if we assign $\Phi_Q^q$ and $\Phi_\Qc^q$ baryon number 
$\pm \frac{1}{3}$ and assign $\Phi_Q^l$ and $\Phi_\Qc^l$ baryon number 
$\mp \frac{2}{3}$. 
We have suppressed proton decay through a missing partner mechanism, 
as long as the colored triplet Higgs boson does not mix quark and lepton
sectors at the GUT scale.

However, the $q$ and $l$ Higgs sectors must not be completely decoupled: 
if this were the case, then only one Higgs pair
would acquire a vev at the electroweak scale, leaving the second
sector massless.  Moreover, we need $\mu$ terms for Higgs fields,
which may be generated by giving a vev to the NMSSM fields
in both Higgs sectors. The trilinear superpotential is 
\begin{eqnarray}
\label{Eq: App NMSSM}
W_{\text{NMSSM}} &=& \frac{1}{3}\kappa_{l} (\Phi_L^l)^3 + \frac{1}{3}\kappa_q 
(\Phi_L^q)^3 +
\half\kappa'_l (\Phi_L^l)^2 \Phi_L^q +\half \kappa'_q (\Phi_L^q)^2 \Phi_L^l + 
\text{ cyclic.}  \\
&\supset&
\left(\begin{array}{cc}h_u &h_\nu\end{array}\right)
 \left( \begin{array}{cc}(\kappa_q n_q + \half \kappa_q' n_l) &
  (\kappa_l' n_l +  \kappa_q' n_q)\\
    (\kappa_l' n_l +  \kappa_q' n_q)& (\kappa_l n_l + \half \kappa_l' n_q)
    \end{array}\right)
\left(\begin{array}{c}h_d\\h_e\end{array}\right) .
 \end{eqnarray}
The low energy theory is similar to the NMSSM, but without $n^{3}$ 
terms\footnote{
Normally, the $n^{3}$ terms explicitly break a Peccei-Quinn (PQ) 
symmetry, preventing the appearance of massless Goldstone mode.  Here, the PQ
symmetry is partially contained within $SU(3)^{3}$, so the
Goldstones are given a mass via a tadpole for $n$.}.
The cyclic permutations of the interactions in Eq.~(\ref{Eq: App NMSSM})
no longer preserve baryon number exactly.  
The $\kappa_l$ and $\kappa_q$ interactions give rise to dimension seven 
$p\rightarrow e^+ \bar{\nu} \bar{\nu}$ and $n \rightarrow \bar{n}$ 
oscillations.  The $\kappa_q'$ and $\kappa_l'$ interactions lead to 
dimension seven proton decay processes of the form $p \rightarrow K^+ \nu$.   
All of these come with three powers of Yukawa couplings, and three
powers of the GUT scale in the denominator. 
 
For example, consider the baryon number violating operator 

\begin{equation}
 W_{\not{B}}=  \frac{qqq  q d^c e^c}{M^{3}} +\cdots,
 \end{equation}
which is generated by the diagram
in Fig. 7.  The interactions in this diagram
come from the cyclic permutations of the terms in $W_{Y}$ and $W_{\text{NMSSM}}$.
Taking into account the small Yukawa couplings and CKM mixings, 
this leads to a proton lifetime 
\begin{equation}
\label{eqn:pdecay}
\tau_{p} \sim 10^{70} \text{yrs.} \left( \frac{M_{H_C}}{M_{\text{GUT}}} \right)^{6},
\end{equation}
where $M_{H_{C}}$ is the mass of the colored Higgs triplet 
mediating the proton decay.  
We can therefore take the mass of the Higgs triplet to be quite low
without leading to unacceptable proton decay, unlike in the MSSM.

\begin{figure}
\begin{center}
\epsfig{file=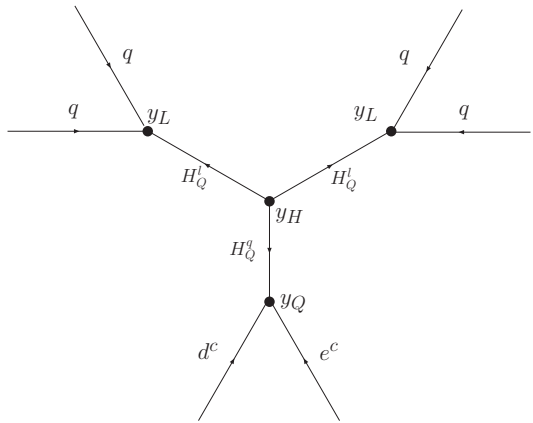,width=2.8in}
\label{fig:Bviolation}
\caption{A leading operator contributing to baryon number violation.}
\end{center}
\end{figure}

\subsection{Precision Unification}
\label{Sec: A Precision Unification}

In Sec.~\ref{Sec: P Precision Unification}, we showed that the effects of 
higher loop running are encoded in the anomalous dimensions of 
the light fields.  In accelerated unification models, 
we must also include contributions from the new light 
fields.   In addition to the second pair of Higgs doublets, 
there are also pseudo-Goldstone bosons (PGBs), 
as we will now discuss.  The
mass of these particles is the largest source of 
imprecision in the gauge coupling unification prediction. 

The superpotential of Eq.~(\ref{Eq: SymSuperpotential}) has a 
global $SU(3)$ symmetry.  The breaking 
to the diagonal will in general give rise to Goldstone bosons.  In the
$SU(3)$ color sector, all such particles get a mass from gauge interactions.  
For the $SU(3)_{L}$ and $SU(3)_R$ sectors, this is not the case.
The condition to lift these PGBs is $\det \mu_I \ne 0$.  
While renormalization group evolution 
splits the $\mu$ parameters, it will not cause the determinant to flow 
to a non-zero value, although GUT-scale threshold corrections could make
this determinant non-zero.  The mass
of the PGBs will then RG evolve as $\det \mu$ evolves.
If the $SU(3)$ breaking couplings are larger than the 
$SU(3)$ preserving superpotential couplings, then $\det \mu$ 
will flow away from zero.  We will not specify the mass of these particles,
but will summarize their possible effect on unification at the end of 
this section.

We now apply the formalism of Appendix A to accelerated unification.
The difference between the $SU(3)$ and $SU(2)$ couplings is
\begin{eqnarray}
\nonumber
\Delta^{32}\equiv \frac{8\pi^2}{g_3^2(\mu)} - \frac{8\pi^2}{g_2^2(\mu)} &=& 
\left[\Delta b^{32}_{\text{MSSM}}\log \frac{\Lambda}{\mu} + \Delta 
b^{32}_{\text{AU}} \log \frac{\Lambda}{\langle\Sigma\rangle} \right]\\ 
\nonumber&&
+ \delta z_{\psi}^{32} +  \delta z_{h}^{32} 
+ 3 \log \frac{g^2_A(\Lambda) g^2_B(\Lambda)}{g^2_3(\mu)}
- 2 \log \frac{g^2_A(\Lambda) g^2_B(\Lambda)}{g^2_2(\mu)} \\
 &&
 +  \delta b^{32}_h\log\frac{\langle \Sigma\rangle}{\mu_{H}}  + \delta 
b^{32}_{\PGB_L}\log\frac{\langle \Sigma\rangle}{m_{\PGB_L}}, 
\label{Eq: PrecisionA32}
\end{eqnarray}
where $\mu_{H}$ is the mass of the second pair of Higgs doublets,
$\Delta b^{32}_{\text{MSSM}}$ is the difference between beta functions
in the MSSM, and 
$\Delta b^{32}_{\text{AU}} \equiv \Delta b^{32}_A+ \Delta b^{32}_B - \Delta 
b^{32}_{\text{MSSM}}$
is the difference between beta function coefficients 
for the additional accelerated unification fields.  When $\NN=2$, 
\mbox{$\Delta b^{32}_{\text{AU}}=\Delta b^{32}_{\text{MSSM}}= -4$.}   
The quantity in brackets reproduces the MSSM one loop prediction. 
The $\log Z$'s of the 
light fields
\begin{eqnarray}
\delta z_\psi^{32} = \half \log \prod_{f=1}^{3} \frac{ Z_{u^c_f} Z_{d^c_f}}{Z_{q_f} 
Z_{l_f}},
\hspace{0.4in}
\delta z_h^{32} = -\half \log  Z_{h_u} Z_{h_d}Z_{h_e} Z_{h_\nu},
\end{eqnarray}
are evaluated by integrating the anomalous dimensions from 
$\Lambda$ down to $\mu$.  The final terms arise from the second 
pair of Higgs doublets and the PGBs in the left sector, with
\begin{equation}
\delta b^{32}_{h} =-1 \hspace{0.5in} \delta b^{32}_{\PGB_L}=  -1. 
\end{equation}
If the masses of these particles were precisely 
at the scale of diagonal breaking, they would not contribute 
any additional deviation.  However, in the model presented here the 
masses are essentially free parameters.  
The mass of the
extra Higgs multiplet depends sensitively on the values of the various
$\mu$ parameters.  A similar calculation 
for the $SU(2)$ and $U(1)$ couplings yields, 
\begin{eqnarray}
\nonumber
\Delta^{21} \equiv \frac{8\pi^2}{g_2^2(\mu)} - \frac{8\pi^2}{g_1^2(\mu)} &=& 
\left[\Delta b^{21}_{\text{MSSM}}\log \frac{\Lambda}{\mu} + \Delta 
b^{21}_{\text{AU}} \log \frac{\Lambda}{\langle\Sigma\rangle} \right] \\ 
&+& 2 \log \frac{g^2_A(\Lambda) g^2_B(\Lambda)}{g^2_2(\mu)}
 + \delta z_{\psi}^{21} +  \delta z_{h}^{21} 
 +  \delta b^{21}_h \log\frac{\langle \Sigma\rangle}{\mu_{H}} \\
&+& \delta b^{21}_{\PGB_L} \log\frac{\langle \Sigma\rangle}{m_{\PGB_L}} 
+ \delta b^{21}_{\PGB_R} \log\frac{\langle \Sigma\rangle}{m_{\PGB_R}}, 
\end{eqnarray}
with $\Delta b^{21}_{\text{MSSM}}=\Delta b^{21}_{\text{AU}}=-18/5$.
The $\log Z$s of the light fields are
\begin{eqnarray}
\delta z_\psi^{21} = \frac{1}{5} \log \Pi_f \frac{ Z_{q_f}^7 Z_{l_f}}{Z_{u^c_f}^4 
Z_{d^c_f} Z_{e^c}^3},
\hspace{0.4in}
\delta z_h^{21} = \frac{1}{5} \log  Z_{h_u} Z_{h_d}Z_{h_e} Z_{h_\nu}.
\end{eqnarray}
Finally, the additional Higgs doublets and the PGBs
contribute to the beta function coefficients
\begin{eqnarray}
\delta b^{21}_h= \frac{2}{5},
\hspace{0.5in} 
\delta b^{21}_{\PGB_L}=  \frac{2}{5} ,
\hspace{0.5in}
\delta b^{21}_{\PGB_R}= - \frac{12}{5} .
\end{eqnarray}
For the moment, let us assume that these additional light fields 
are degenerate with the remainder of the multiplet.
In this case, we find the ratio 
\begin{eqnarray}
\frac{\Delta^{32}}{\Delta^{21}} \simeq 0.721.
\end{eqnarray}
This result is insensitive to the choice of ${\alpha_{A}^{-1}}$.  
Making the more reasonable choice that the PGBs and the extra Higgs 
multiplet are two $e$-folds below 
the rest of the multiplet, we find $\Delta^{32}/\Delta^{21} = 0.716$. 
Again, we conclude that none of the calculable corrections to unifications 
are very large.  
However, if the masses deviate too far from the diagonal breaking scale
the corrections from the PGBs can be non-negligible.
Moreover, when $\NN > 2$, there are more PGBs, which can amplify these
effects.

There are of course additional threshold corrections.  As before, there
are the corrections from SUSY breaking. Also, since 
there are two copies of $SU(3)^3$
near the GUT scale, the high energy particle content is double that
of the MSSM.  So, the naive expectation for GUT scale threshold corrections is 
that they should be roughly double those of the MSSM.
An interesting possibility arises in accelerated unification that is 
not present in the MSSM.  In Sec.~\ref{Sec: P Decay} we showed
that proton decay is suppressed, so it is possible for 
the colored Higgs triplets that mediate proton decay to be much lighter 
than the GUT scale.  This leads to a threshold correction that improves 
the  unification of the couplings.  In the minimal SU(5) GUT, such a 
threshold correction would be desirable, but is forbidden by proton decay 
\cite{Murayama:2001ur}.

\section{Conclusions} 

Gauge coupling unification
places a natural constraint on the structure of potential gauge 
extensions of the MSSM. 
Moreover, it limits the size of the new gauge couplings under which the Higgs
boson may be charged.  The result is that the Higgs mass cannot be too heavy, 
even in models with extended gauge structure.  Accelerated unification
seems to be the best hope for realizing a heavier Higgs mass, but due to 
the presence of pseudo-Goldstone bosons, unification becomes somewhat delicate.

In the models discussed here, a host of new states associated with the 
breaking should be found at the 3-10 TeV scale.  In the accelerated 
unification model, it is likely that the lightest state would be one of 
the PGBs.  The precise mass of this particle depends on the breaking of 
the GUT symmetry.  We now discuss precision electroweak constraints, which 
set the mass of the new vector bosons.

\subsection{Constraints}

Precision electroweak constraints on these models arise from interactions
between the heavy vector bosons and the Standard Model fermions and Higgs.
One might worry that these considerations significantly constrain
the theory; in order to maximize non-decoupling $D$-terms we
need the Standard Model gauge couplings to be as strong as possible, which
means that the heavy vector bosons couple with $\OO(1)$ strength to the
Standard Model fermions.  However, because these new vectors are not
responsible for cutting off the gauge quadratic divergences to the Higgs, they
can be quite heavy without fine-tuning.  
Instead, the heavy vectors cut off divergent contributions to
the Higgs mass arising from the modified quartic coupling, so
the vectors may comfortably lie in the 3 -- 10 TeV range.  

To see this, it is useful to introduce mixing angles $\theta_i$ 
for the gauge fields, obeying
\begin{eqnarray}
\sin^2 \theta_i =\frac{g^2_{i,\SM}}{g^2_{i,A}}, 
\hspace{0.3in} \cos^2 \theta_i =\frac{g^2_{i,\SM}}{g^2_{i,B}}.
\end{eqnarray}
The heavy vectors, which we denote $A_i$, 
couple to the to the MSSM via the interaction
\begin{eqnarray}
\LL_{\eff}= -  g_{i,\SM} \cot \theta_i\, A_i^\mu j_{\mu i, \SM} + 
\frac{g_{i,SM}^2f^2}{\sin 2\theta_i} (A_i^\mu)^2.
\end{eqnarray}
At low energies 
the $A_i$  may be integrated out to give the current-current interaction
\begin{eqnarray}
\LL_{\text{eff}} =  \frac{\cos^4\theta_i}{ 2f^2} j^\mu_{i,\SM} 
j_{\mu\text{SM}}
.\end{eqnarray}
These terms contribute to the $W,Y,Z$ extended oblique corrections,
which are constrained experimentally, implying a constraint 
\cite{Barbieri:2004qk}
\begin{eqnarray}
\frac{f_{L,R}}{\cos^2 \theta_{L,R}} \gsim 3.5 \TeV.
\end{eqnarray}
Thus even for $\cos\theta_{L,R}\sim 1$ the breaking scales can be 3.5 TeV and
vectors will be under 10 TeV.  

\subsection{Future Directions}

There are several potential directions for future work. As noted in the text, 
the theories under consideration are very similar to 
deconstructed models of gaugino mediation.
It would be interesting to determine whether the link fields can 
communicate SUSY breaking to the MSSM.  Once a SUSY breaking scenario is 
specified, either this mechanism or another, 
it would be possible to discuss spectroscopy 
and unification in further detail.  It would also be of interest to explore 
the supersymmetric four Higgs doublet in more detail.  In principle, the experimental limits on the Higgs boson can be modified.

\section*{Acknowledgments}
We thank Nima Arkani-Hamed, Puneet Batra, Spencer Chang, Savas Dimopoulos, 
Howie Haber, Shamit Kachru, Ami Katz, Michael Peskin, and Eric Poppitz
for useful discussions.

\appendix
\section{Precision Unification and Holomorphy}
In supersymmetric theories, threshold corrections are constrained by holomorphy. 
This technique can be applied to calculate corrections to unification 
in any model where holomorphy is a useful constraint (i.e. when there are
large supersymmetric masses).

Our first result is that threshold effects from mass splittings 
cancel against higher loop corrections \cite{NSVZ,NimaHitoshi}.  To see 
this, consider the exact NSVZ beta function \cite{NSVZ}
\begin{eqnarray}
\beta_g = \frac{g^3}{16\pi^2} \frac{ b_0 -  \sum_a T_2(a) \gamma_a}{1 - C_2 
\frac{g^2}{8 \pi^2}},
\end{eqnarray}
where $b_0 = -3 C_2 + \sum_a T_2(a)$.  
This can be integrated to give
\begin{eqnarray}
\label{Eq: Int NSVZ}
\frac{8\pi^2}{g^2(\mu)} + C_2 \log g^2(\mu)
= \frac{8\pi^2}{g^2(\Lambda)} + C_2 \log g^2(\Lambda)
+ b_0 \log \frac{\Lambda}{\mu} - \sum_a T_2(a) \log Z_a(\mu, \Lambda).
\end{eqnarray}
Now consider integrating out a massive matter field  (like the link fields, $\Sigma$).  
Gauge couplings are matched at the physical mass of the 
field, $m_r$, which differs from the 
holomorphic mass $m$ (which appears in the superpotential) 
by a factor of the wave function renormalization: $m = Z(m_r, \Lambda) m_r$.  
Thus, the $\log Z$ that appears in the NSVZ formula can be combined with a
holomorphic mass to recover a running mass. So, it is possible to write a 
RG invariant matching equation exclusively in terms of holomorphic quantities:
\begin{eqnarray}
\label{Eq: Holo Match}
\frac{8\pi^2}{g^2_\LE(\Lambda)} + C_2 \log g^2_\LE(\Lambda)= 
\frac{8\pi^2}{g^2_\HE(\Lambda)}+ C_2 \log g^2_\HE(\Lambda) -  \log \frac{\Lambda}{m}.
\end{eqnarray}
Here $g^2_\LE(\Lambda)$ and $g^2_\HE(\Lambda)$ are the low energy 
and high energy gauge couplings defined at the cut-off $\Lambda$; they 
have one loop beta functions $b_{0,\LE}$ and $b_{0,\HE}$ respectively that 
differ by one.   
Using the NSVZ beta function, one can verify 
that Eq.~(\ref{Eq: Holo Match}) is equivalent the matching
the high energy and low energy gauge couplings at the 
physical mass scale.
However, Eq.~(\ref{Eq: Holo Match})
is valid all scales, including at the cut-off,  
where there clearly has been no running to split $m$.
Thus, complete GUT multiplets will lead to a small deviation from the 
MSSM prediction.   The dominant effect is indirect: the gauge 
coupling RG trajectories are deflected by the presence of 
the $\Sigma$ fields, which in turn contributes to the last term 
in Eq.~(\ref{Eq: Int NSVZ}) for the light MSSM fields.

A second potential source of modifications to unification comes from
the breaking of extended gauge symmetry.  We must apply a 
matching condition when $G \times G_{GUT} \rightarrow G_{SM}$.
The usual matching conditions 
\begin{eqnarray}
\frac{8\pi^2}{g^2(m_{V,\phys})} = 
\frac{8\pi^2}{g_A^2(m_{V,\phys})} 
+\frac{8\pi^2}{g_B^2(m_{V,\phys})}, \hspace{0.4in}
\left(\frac{8\pi^2}{g^2(m_{X,\phys})}\right)_{-} = 
\left(\frac{8\pi^2}{g^2(m_{X,\phys})}\right)_{+}, 
\end{eqnarray}
are reproduced by the RG invariant matching equation
\begin{eqnarray}
\label{Eq: P RGInv}
\frac{8\pi^2}{g^2(\Lambda)} + C_2 \log g^2(\Lambda) = 
\frac{8\pi^2}{g_A^2(\Lambda)} + C_2 \log g_A^2(\Lambda)
+\frac{8\pi^2}{g_B^2(\Lambda)} + C_2^{B} \log g_B^2(\Lambda)
- C_2^{B} \log \frac{\Lambda^2}{\langle\Sigma \Sigma^c\rangle}.
\end{eqnarray}
Here $\langle\Sigma\rangle$ and $\langle\Sigma^c\rangle$ are the vevs of the 
fields that break the gauge symmetry, and we have used 
$C_{2}^{SM}=C_{2}^{A} \equiv C_{2}$. We have also
used expressions for the renormalized gauge boson masses:
\begin{eqnarray}
M_{V}^{2} &=& Z(M_{V}, \Lambda)  \langle\Sigma \Sigma^c\rangle (g_{A}^{2} + 
g_{B}^{2})_{M_{V}^{2}} \\
M_{X}^{2} &=& Z(M_{X}, \Lambda)  \langle\Sigma \Sigma^c\rangle 
(g_{B}^{2})_{M_{V}^{2}} .
\end{eqnarray}
Applying the above RG invariant matching condition gives rise 
to Eq.~(\ref{Eq: P 32 Diff}) for product unification.

For accelerated unification, the RG invariant matching equation is similar.
The analog of Eq.~(\ref{Eq: P RGInv}) is:
\begin{eqnarray}
\nonumber
\frac{8\pi^2}{g^2(\mu)} + C_2 \log g^2(\mu) &=& 
\frac{8\pi^2}{g_A^2(\Lambda)} + C_2 \log g_A^2(\Lambda)
+\frac{8\pi^2}{g_B^2(\Lambda)} + C_2 \log g_B^2(\Lambda)\\
&&\nonumber
- C_2 \log \frac{\Lambda^2}{\langle\Sigma \Sigma^c\rangle}
+ b_{\phi} \log \frac{\Lambda}{m_\phi}\\
&&
  + b_{0,\text{light}} \log \frac{\Lambda}{\mu}- \sum_{a} T_2(a)  \log 
Z_a(\mu,\Lambda).
\end{eqnarray}
We have used $C_{2} =C^{A}_{2} =C^{B}_{2}$, and integrated out the $\phi$ multiplet 
at its holomorphic mass.    In this case, there
are additional light states in the sum, namely the pseudo-Goldstone bosons 
and Higgs 
multiplets discussed in the text.  

\section{D-Terms and Tadpoles}
In this appendix, we give general expressions for the masses 
of $M_{V_{H}}$, $S$, $\phi$, and $\eta$, taking into account the possibility 
of a SUSY breaking induced tadpole for the $S$ field. 
After including the supersymmetry breaking, as in Eq.~(\ref{Eq:spurion}),
and restricting attention to the field $S$, there is a linear source term: 
\begin{eqnarray}
\nonumber
W&=&
((a-b)\mu- m^2)\theta^2 \sqrt{6} \frac{\mu}{\lambda} S
- \mu(1 + (2a-b)\theta^2 ) \frac{S^2}{2}
+ \frac{\lambda}{3\sqrt{6}}(1+a\theta^2)S^3 \\
K&=&  (1+ m^2\theta^4) S^\dagger S .
\end{eqnarray}
We may shift $S$  by a constant to remove this term,
\begin{eqnarray}
 S \rightarrow S+ J \hspace{0.5in} J=\frac{\sqrt{6}}{\lambda}(-j_S\mu +  
j_F\mu^2\theta^2) .
\end{eqnarray}
Solving for $j_S$ and $j_F$,
\begin{eqnarray}
\label{Eq: Source Eq1}
&&j_F = -j_S -  j_S^2\\
&&\frac{ \mu^2 - m^2 + (2a - b)\mu}{\mu^2} j_S
+ \Big( 3 + \frac{a}{\mu}\Big) j_S^2
+ 2  j_S^3=\frac{(m^2 + (b-a)\mu)}{\mu^2}.
\end{eqnarray}
This shift affects the Kahler term for $V_H$: 
\begin{eqnarray}
K= g_H^2\frac{\mu^2}{\lambda^2}\left( 1+ j_S\right)^2
\left(1+ \left(m^2+ \mu^2 \frac{j_F^2}{(1+ j_S)^2}\right)\theta^4\right)
 \Tr V_H^2 .
\end{eqnarray}
This expression summarizes how SUSY breaking is communicated to the vector 
multiplet.  
We must also check that the other scalar masses
\begin{eqnarray}
\nonumber
\frac{m_S^2}{\mu^2} &=& \Big( 1 + 2 j_S\Big)^2 + \frac{m^2}{\mu^2}
\pm  \Big( \frac{2a - b}{\mu} -2 \Big( j_F - \frac{a}{\mu} j_S\Big) \Big)\\ 
\nonumber
\frac{m_\phi^2}{\mu^2} &=& \Big( 2 +  j_S\Big)^2 + \frac{m^2}{\mu^2}
\pm  \Big( \frac{a+ b}{\mu} - \Big( j_F - \frac{a}{\mu} j_S\Big) \Big)\\ 
\frac{m_\eta^2}{\mu^2} &=& \left( (3 + 2j_S)^{2} + 2 j_{S}^{2} +\frac{m^{2}}{\mu^{2}} \right) \mp
3\left[4 j_S^2 -5 j_{S} -\frac{m^{2}}{\mu^{2}} + (1+ j_{S}) \frac{a}{\mu} \right]
\end{eqnarray}
remain positive.  
It may be shown that these masses remain positive over large regions of
parameter space.
However, as described in the text, it is sufficient to note that when
$a=0$ and $j_{S} = j_{F} =0$ the masses are positive whenever 
$m^{2} > -\mu^{2}/2$.


\begin{thebibliography}{99}

\bibitem{Barate:2003sz}
R.~Barate {\it et al.}  [ALEPH Collaboration],
Phys.\ Lett.\ B {\bf 565}, 61 (2003)
[arXiv:hep-ex/0306033].

\bibitem{Kane:2004tk}
G.~L.~Kane, B.~D.~Nelson, L.~T.~Wang and T.~T.~Wang,
arXiv:hep-ph/0407001.

\bibitem{DEKPuneet}
P.~Batra, A.~Delgado, D.~E.~Kaplan and T.~M.~P.~Tait,
JHEP {\bf 0402}, 043 (2004) [arXiv:hep-ph/0309149].

\bibitem{DEKPuneet2}
P.~Batra, A.~Delgado, D.~E.~Kaplan and T.~M.~P.~Tait,
JHEP {\bf 0406}, 032 (2004)
[arXiv:hep-ph/0404251].

\bibitem{Harnik:2003rs}
R.~Harnik, G.~D.~Kribs, D.~T.~Larson and H.~Murayama,
arXiv:hep-ph/0311349;
S.~Chang, C.~Kilic and R.~Mahbubani,
arXiv:hep-ph/0405267.

\bibitem{Birkedal:2004xi}
A.~Birkedal, Z.~Chacko and M.~K.~Gaillard,
arXiv:hep-ph/0404197.

\bibitem{Polonsky}
N.~Polonsky and S.~Su,
Phys.\ Lett.\ B {\bf 508}, 103 (2001)
[arXiv:hep-ph/0010113];
P.~Langacker, N.~Polonsky and J.~Wang,
Phys.\ Rev.\ D {\bf 60}, 115005 (1999)
[arXiv:hep-ph/9905252].


\bibitem{Savas}
S.~Dimopoulos and H.~Georgi,
Nucl.\ Phys.\ B {\bf 193}, 150 (1981);
S.~Dimopoulos, S.~Raby and F.~Wilczek,
Phys.\ Rev.\ D {\bf 24}, 1681 (1981).

\bibitem{LR}
L.~Randall.  Talk at SUSY 2002.

\bibitem{Arkani-Hamed:2001vr}
N.~Arkani-Hamed, A.~G.~Cohen and H.~Georgi,
arXiv:hep-th/0108089.

\bibitem{MissingPartner}
B.~Grinstein,
Nucl.\ Phys.\ B {\bf 206}, 387 (1982).
%
A.~Masiero, D.~V.~Nanopoulos, K.~Tamvakis and T.~Yanagida,
Phys.\ Lett.\ B {\bf 115}, 380 (1982).

\bibitem{NMSSM}
P.~Fayet,
Nucl.\ Phys.\ B {\bf 113}, 135 (1976);
H.~P.~Nilles, M.~Srednicki and D.~Wyler,
Phys.\ Lett.\ B {\bf 120}, 346 (1983).
J.~P.~Derendinger and C.~A.~Savoy,
Nucl.\ Phys.\ B {\bf 237}, 307 (1984).
J.~R.~Ellis, J.~F.~Gunion, H.~E.~Haber, L.~Roszkowski and F.~Zwirner,
Phys.\ Rev.\ D {\bf 39}, 844 (1989).

\bibitem{MSSMHiggsBound}
J.~R.~Ellis, G.~Ridolfi and F.~Zwirner,
Phys.\ Lett.\ B {\bf 257}, 83 (1991); 
Y.~Okada, M.~Yamaguchi and T.~Yanagida,
Prog.\ Theor.\ Phys.\  {\bf 85}, 1 (1991); 
H.~E.~Haber and R.~Hempfling,
Phys.\ Rev.\ Lett.\  {\bf 66}, 1815 (1991).

\bibitem{Espinosa}
J.~R.~Espinosa and M.~Quiros,
Phys.\ Rev.\ Lett.\  {\bf 81}, 516 (1998)
[arXiv:hep-ph/9804235].

\bibitem{KaneKolda}
G.~L.~Kane, C.~F.~Kolda and J.~D.~Wells,
Phys.\ Rev.\ Lett.\  {\bf 70}, 2686 (1993)
[arXiv:hep-ph/9210242].

\bibitem{PDG}
S.~Eidelman {\it et al.}  [Particle Data Group Collaboration],
Phys.\ Lett.\ B {\bf 592}, 1 (2004).

\bibitem{Deconstruction}
N.~Arkani-Hamed, A.~G.~Cohen and H.~Georgi,
Phys.\ Rev.\ Lett.\  {\bf 86}, 4757 (2001)
[arXiv:hep-th/0104005].

\bibitem{DeconGauge}
C.~Csaki, J.~Erlich, C.~Grojean and G.~D.~Kribs,
Phys.\ Rev.\ D {\bf 65}, 015003 (2002)
[arXiv:hep-ph/0106044]; 
H.~C.~Cheng, D.~E.~Kaplan, M.~Schmaltz and W.~Skiba,
Phys.\ Lett.\ B {\bf 515}, 395 (2001)
[arXiv:hep-ph/0106098].

\bibitem{Transparent}
D.~E.~Kaplan, G.~D.~Kribs and M.~Schmaltz,
Phys.\ Rev.\ D {\bf 62}, 035010 (2000)
[arXiv:hep-ph/9911293].

\bibitem{Trinified}
S.~L.~Glashow,
Print-84-0577 (BOSTON);
G.~Lazarides, C.~Panagiotakopoulos and Q.~Shafi,
Phys.\ Lett.\ B {\bf 315}, 325 (1993)
[Erratum-ibid.\ B {\bf 317}, 661 (1993)]
[arXiv:hep-ph/9306332].

\bibitem{Drees:1988fc}
M.~Drees,
Int.\ J.\ Mod.\ Phys.\ A {\bf 4}, 3635 (1989).

\bibitem{NSVZ}
M.~A.~Shifman and A.~I.~Vainshtein,
Nucl.\ Phys.\ B {\bf 277}, 456 (1986)
[Sov.\ Phys.\ JETP {\bf 64}, 428 (1986\ ZETFA,91,723-744.1986)].
%
V.~A.~Novikov, M.~A.~Shifman, A.~I.~Vainshtein and V.~I.~Zakharov,
Nucl.\ Phys.\ B {\bf 229}, 381 (1983).
%
V.~A.~Novikov, M.~A.~Shifman, A.~I.~Vainshtein and V.~I.~Zakharov,
Nucl.\ Phys.\ B {\bf 260}, 157 (1985)
[Yad.\ Fiz.\  {\bf 42}, 1499 (1985)].
%
M.~A.~Shifman, A.~I.~Vainshtein and V.~I.~Zakharov,
Phys.\ Lett.\ B {\bf 166}, 334 (1986).

\bibitem{NimaHitoshi}
N.~Arkani-Hamed and H.~Murayama,
Phys.\ Rev.\ D {\bf 57}, 6638 (1998)
[arXiv:hep-th/9705189].
%
N.~Arkani-Hamed and H.~Murayama,
JHEP {\bf 0006}, 030 (2000)
[arXiv:hep-th/9707133].

\bibitem{Linear Term}
J.~Bagger and E.~Poppitz,
Phys.\ Rev.\ Lett.\  {\bf 71}, 2380 (1993)
[arXiv:hep-ph/9307317].
%
J.~Bagger, E.~Poppitz and L.~Randall,
Nucl.\ Phys.\ B {\bf 455}, 59 (1995)
[arXiv:hep-ph/9505244].
V.~Jain,
Phys.\ Lett.\ B {\bf 351}, 481 (1995)
[arXiv:hep-ph/9407382].
%
C.~Panagiotakopoulos and K.~Tamvakis,
Phys.\ Lett.\ B {\bf 469} (1999) 145
[arXiv:hep-ph/9908351].
%
C.~Panagiotakopoulos and A.~Pilaftsis,
Phys.\ Rev.\ D {\bf 63}, 055003 (2001)
[arXiv:hep-ph/0008268].
%
A.~Dedes, C.~Hugonie, S.~Moretti and K.~Tamvakis,
Phys.\ Rev.\ D {\bf 63}, 055009 (2001)
[arXiv:hep-ph/0009125].

\bibitem{Tuhin}
T.~Roy,
arXiv:hep-ph/0408291.

\bibitem{Murayama:2001ur}
H.~Murayama and A.~Pierce,
Phys.\ Rev.\ D {\bf 65}, 055009 (2002)
[arXiv:hep-ph/0108104].

\bibitem{Barbieri:2004qk}
R.~Barbieri, A.~Pomarol, R.~Rattazzi and A.~Strumia,
arXiv:hep-ph/0405040.

\end{thebibliography}
\end{document}